\documentclass[12pt, preprint]{aastex}

\def\ld{L_{\rm dust}/L_{\star}}
\def\fd{F_{\rm dust}/F_{\star}}

\def\MEarth{M_\oplus}

\usepackage{graphicx}
\graphicspath{{converted_graphics/}}
\begin{document}

\title{Explorations Beyond the Snow Line: {\it Spitzer}/IRS Spectra of Debris Disks Around Solar-Type Stars}

\author{
S.~M. Lawler$^{1,2,3}$,
C.~A. Beichman$^{2}$, 
G. Bryden$^{4}$, 
D.~R. Ciardi$^{2}$, 
A.~M. Tanner$^{4}$,
K.~Y.~L. Su$^{5}$,
K.~R. Stapelfeldt$^{4}$,
C.~M. Lisse$^{6}$,
D.~E. Harker$^{7}$
}
\affil{1) Astronomy Department, Wesleyan University, Middletown, CT 06459}
\affil{2) NASA Exoplanet Science Institute, California Institute of Technology, 
 Pasadena, CA 91125} 
\affil{3) present address: University of British Columbia, Department of Physics and Astronomy, 6244 Agricultural Road, Vancouver, BC V6T 1Z1 Canada}
\affil{4) Jet Propulsion Laboratory, 4800 Oak Grove Drive, Pasadena, CA 91109} 
\affil{5) Steward Observatory, University of Arizona, 933 North Cherry Avenue, Tucson, AZ 85721}
\affil{6) Johns Hopkins University - Applied Physics Laboratory, SD/SRE, MP3-W155, 7707 Montpelier Road, Laurel, MD 20723}
\affil{7) Center for Astrophysics and Space Sciences, University of California, San Diego, 9500 Gilman Drive, La Jolla, CA 92093-0424}

\shorttitle{IRS Spectra of Solar-Type Stars}
\shortauthors{Lawler et al.}

\begin{abstract}

We have observed 152 nearby solar-type stars with the Infrared Spectrometer (IRS) on the \emph{Spitzer
Space Telescope}. Including stars that met our criteria but were observed in other surveys, we get an overall success rate for finding excesses in the
long wavelength IRS band (30--34 $\mu$m) of  11.8$\%\pm2.4\%$.  The success rate for
excesses in the short wavelength band (8.5--12 $\micron$) is $\sim$1$\%$ including
sources from other surveys. For stars with no excess at 8.5--12~$\mu$m, the IRS data set $3\sigma$ limits of
around 1,000 times the level of zodiacal emission present in our
solar system, while at 30--34~$\mu$m set limits of around 100 times the
level of our solar system. Two stars (HD~40136 and HD~10647)  show
weak  evidence for spectral features; the excess emission in
the other systems is featureless. If the emitting material consists 
of large (10~$\mu$m) grains as implied by the lack of spectral features, we find that these grains are typically located at or beyond the snow line,
$\sim$1--35~AU from the host stars, with an average distance of 14 $\pm$ 6~AU; however smaller grains could be located at
significantly greater distances from the host stars. These distances correspond to dust temperatures in the range $\sim$50--450~K.  Several of the disks are well modeled by a single dust temperature, possibly indicative of a ring-like structure.  However, a single dust temperature does not match the data for other disks in the sample, implying a distribution of temperatures within these disks.  For most stars with excesses, we detect an excess at both IRS and MIPS wavelengths.  Only three stars in this sample show a MIPS 70 $\micron$ excess with no IRS excess, implying that very cold dust is rare around solar-type stars.  

\end{abstract}

\keywords{infrared: stars --- circumstellar matter
--- planetary systems --- Kuiper Belt}

%%%%%%%%%%%%%%%%%%%%%%%%%%%%%%%%%%%%%%%
\section{Introduction}

Mid-infrared spectroscopic observations of some young debris stars
such as $\beta$ Pictoris \citep{telesco91}, 51 Oph \citep{fajardo93},
and BD+20 307 \citep[HIP 8920;][]{song05} have revealed warm dust
composed, at least in part, of small (sub-micron) grains of
crystalline silicates such as forsterite and enstatite. The
similarity of these spectral features to those seen in comet C/1995
O1 \citep[Hale-Bopp; e.g.,][]{wooden00} suggests that this
circumstellar material may represent debris from either cometary or
asteroidal material located within the habitable zones of the stars.
Dramatically, observations with the Infrared Spectrometer on the \emph{Spitzer Space Telescope} \citep[IRS;][]{houck04} revealed a bright spectrum of features due to hot (400 K)
silicate grains around the nearby (12.6 pc), mature (2 Gyr) K0 V
star, HD 69830 \citep{beichman69830}. This star, with a level of
exo-zodiacal emission $\sim$1,400 times that of our own solar system, is
also accompanied by a trio of Neptune-mass planets which may be trapping
material in an exterior 2:1 resonance at $\sim$1 AU \citep{lisse07}.
However, these spectral features are not present in all stars with
debris disks. More than a dozen classic debris disks, around mostly
mature stars (including Fomalhaut), examined by \emph{Spitzer}
\citep{jura04, stapelfeldt04} show little or no spectral structure
while showing clear excess at these wavelengths. Similarly, most of
the other stars with excesses in other surveys with IRS show no
evidence for small grains, suggesting that the grains in these
systems are larger than $\sim$10 $\mu$m \citep{beichman06IRS,
chen06}. These grains may be similar to those in our own zodiacal
cloud which are predominantly larger than 10-100 $\mu$m with some
smaller silicate grains, yielding only a weak 10 $\mu$m emission
feature \citep[e.g.,][]{reach03}.

We have used IRS on \emph{Spitzer} to observe a sample of FGKM stars
within 25 pc of the Sun to assess the frequency, amount, and
properties of the warm dust located within the habitable zones
around solar-like stars. Some stars also have data from the Multiband Imaging Photometer for \emph{Spitzer} \citep[MIPS;][]{rieke04}, providing additional information about cool dust located in the Kuiper belts of these systems.  This information can shed light on the
formation and evolution of circumstellar material located relatively
close to the host star. 

This study addresses the nature of
asteroidal and cometary material, which as the techniques of planet
detection improve, may prove to be tracers for gas giant and rocky planets.  This is highlighted by the discovery of three planets orbiting in the immediate vicinity of the HD 69830 debris disk \citep{lovis06}, as well as by the recent images of an exoplanet within the annulus of Fomalhaut's debris disk \citep{kalas08}, and three exoplanets around HR 8799 \citep{marois08}, which was previously known to have an IR excess \citep{rhee07}.  
Together with planets, this circumstellar material forms complete planetary systems \citep{beichmanPPV}.  In this paper we discuss our sample selection ($\S$ \ref{sample}); review our reduction procedure and present our spectra ($\S$ \ref{obsdata}); discuss measured IR excesses ($\S$ \ref{results}); present our models and discuss the nature of the debris disks around 19 stars with detected IRS and/or MIPS 70 $\micron$ excesses ($\S$ \ref{discussion}); and review implications of our results for debris disks around solar-type stars ($\S$ \ref{conclusion}).  

%%%%%%%%%%%%%%%%%%%%%%%%%%%%%%%%%%%%
\section{The Sample} \label{sample}

The primary goal of our IRS survey is to perform a uniform census of
nearby FGKM stars to determine the frequency and amount of warm dust
located within the habitable zones of these stars. The survey 
complements our more complete understanding of the frequency and
amount of the cold dust located near the Kuiper Belts of solar-type
stars \citep[e.g.,][]{bryden06}.  We have chosen a sample of
solar-like stars (spectral types F, G, K and early M) from the Hipparcos dataset  based upon the
following criteria: a) effective temperature in the range 7300 $\gtrsim$ T$_{eff}$ $\gtrsim$ 3800 K
corresponding to F0--M0 spectral types; b) luminosity class V; c) distance within 25 pc of the Sun; d) no  nearby stellar companions; e) not  variable as identified  by Hipparcos or other catalogs; f) predicted $F_\nu(30 \mu$m) flux density of at least 30 mJy; g) not observed previously by \emph{Spitzer} with IRS as of 2004
when this sample was defined. This last criterion eliminated 51
stars of which 8 have IRS excesses; 
these numbers have been taken into account in the statistics of detections discussed in $\S$ \ref{statdetect}.  This sample does not include every star that meets these criteria, but stars in the sample have been chosen somewhat randomly, so this should represent an unbiased sample of stars meeting these criteria.

\begin{figure}
\includegraphics[scale=0.7]{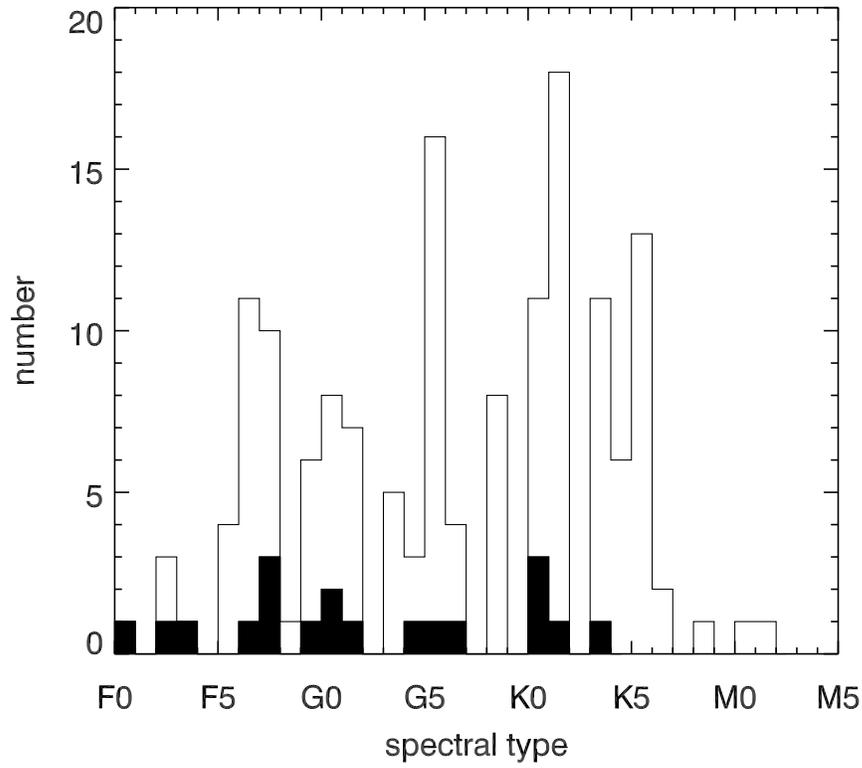}
\caption{Spectral types of the sample stars. Stars with excesses are
noted by filled bars. \label{sptypehists}}
\end{figure}

\begin{figure}
\includegraphics[scale=0.7]{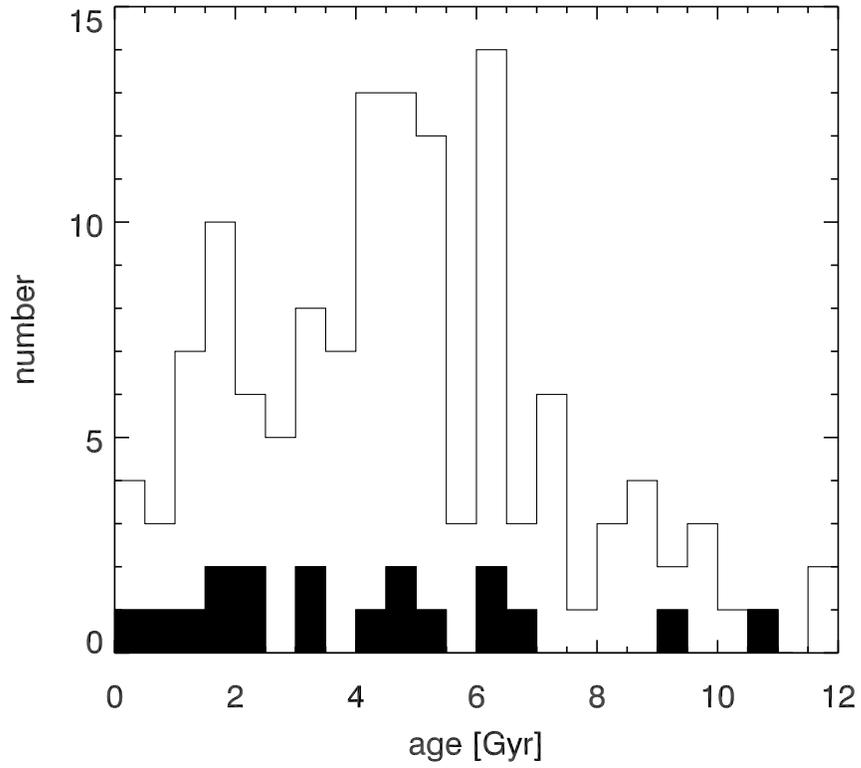}
\caption{Ages of the sample stars. Stars with excesses are noted by
filled bars. Ages are from \citet{wright04} or \citet{valenti05} if available, otherwise we use an average of
literature values (see Table~\ref{basictable}).
\label{agehists}}
\end{figure}

\begin{figure}
\includegraphics[scale=0.7]{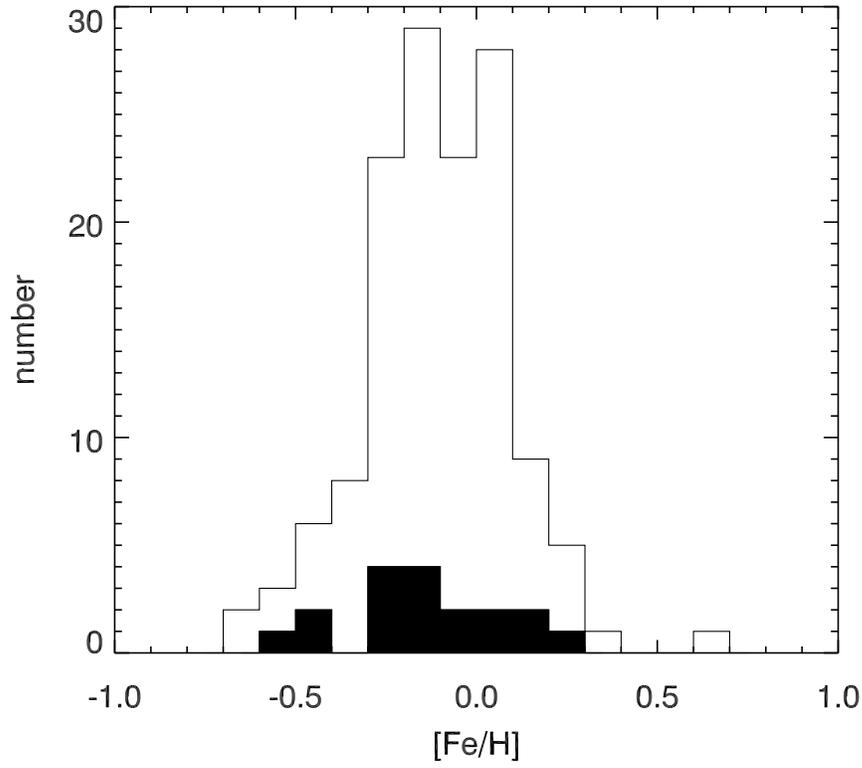}
\caption{Metallicities of the sample stars. Stars with excesses are
noted by filled bars. We use an average of literature values (see
Table~\ref{basictable}). 
\label{methists}}
\end{figure}

There are 152 stars in the sample, distributed fairly evenly in
spectral type. The ends of the distribution are not as well
populated, mostly as a result of the distance criterion at the
bright end, and the minimum flux density criterion at the faint end.
Figures~\ref{sptypehists}--\ref{methists} show the distribution of
stars in spectral type, age, and metallicity, which are listed for
each star in Table \ref{basictable}.

The most uncertain stellar parameter is, of course, age, for these
mature, main sequence stars. While the values given in Table
\ref{basictable} (and shown in Figure~\ref{agehists})
are derived from many heterogeneous sources, we gave priority to
spectroscopic determinations from \citet{wright04} or \citet{valenti05}. If not
from these two sources, quoted values are an average of a wide
variety of values taken from the literature. Thus, the age of any
given star must be regarded with caution, i.e.,\ not more
accurate than a factor of two. Of the sample overall, it is safe to
say that the vast majority are older than 1 Gyr, well beyond the age
when infrared excesses are known to be common among A--G stars
\citep{su06, siegler07}. HD 10647 highlights the problems with
determining ages. \citet{chen06} suggest 300 Myr, and the common space motion of this star with the Tucanae-Horlogium association lends credence to a young age estimate \citep{zuckerman04}. However, \citet{valenti05} suggest an age
around 2--4 Gyr. We use the younger value of 300 Myr.

%****asdf
Of the 152 stars selected, we have MIPS data at 24 $\mu$m and 70 $\mu$m for 78 stars from a variety of programs (noted in Table~\ref{IRStable}).  
%We have MIPS data at 24 $\mu$m and 70 $\mu$m for all 152 stars from a variety of programs (noted in Table~\ref{IRStable}).  
These data permit us to
cross-correlate the longer wavelength detections of cooler, Kuiper
Belt dust with our shorter wavelength detections of hotter dust,
yielding a more complete understanding of the dust distribution and
mass within exo-zodiacal clouds.

%%%%%%%%%%%%%%%%%%%%%%%%%%%%%%%%%%%%
\section{Observations and Data Reduction} \label{obsdata}

We observed each star with all four wavelength modules of the IRS:
Short-Low Order 2 and 3 (SL2; 5.1--7.5 $\micron$, SL3; 7.1--8.4
$\micron$), Short-Low Order 1 (SL1; 7.5--14.0 $\micron$), Long-Low
Order 2 (LL2; 14.0--20.5 $\micron$), and Long-Low Order 1 (LL1; 20--34
$\micron$), as part of the {\it Spitzer} GO program 20463 (D. Ciardi,
P.I.). The basic observing sequence and associated data reduction have been described in
\citet{beichman05FGK} and \citet{beichman06IRS}. In summary, we
have used the fact that the vast majority of the sample ($>$85\%)
shows no excess in an initial examination of the IRS data or in
longer wavelength MIPS data to derive a ``superflat'' to improve the
relative calibration of all the spectra and thus to make small
deviations from expected photospheric levels detectable with the
greatest possible sensitivity.

\begin{figure}
\includegraphics[scale=0.7]{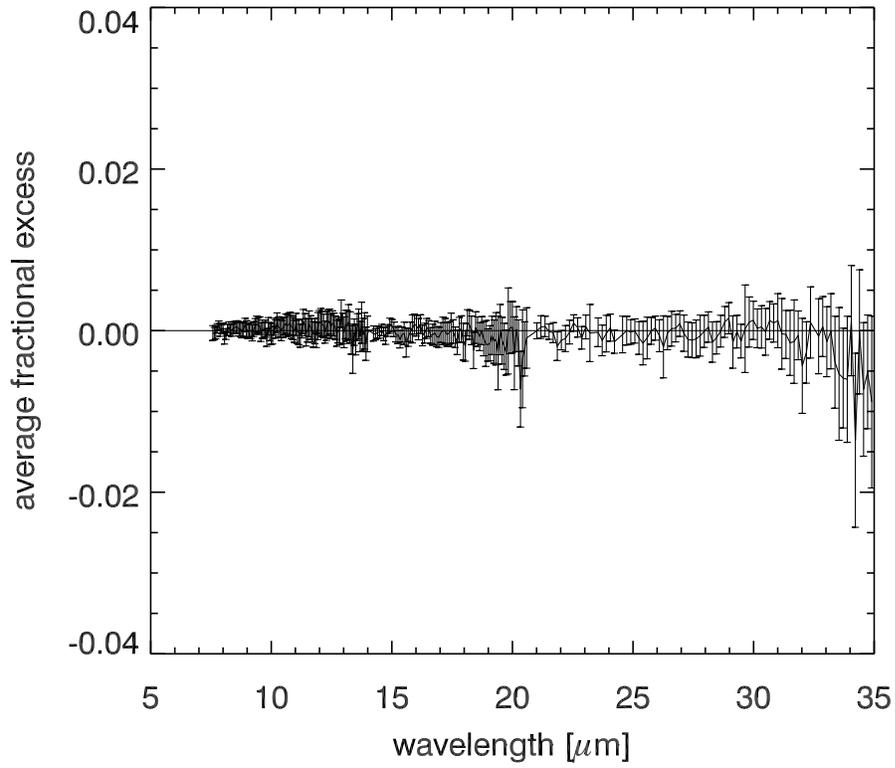}
\caption{Average fractional excess for 126 stars with no excesses averaged by
wavelength. These spectra are beautifully calibrated, with an
average deviation from the photosphere of less than 0.5$\%$.
\label{fracex_noex}}
\end{figure}

The data reduction procedure started  with the Spitzer Science
Center (SSC)-calibrated spectrum, obtained either from images
resulting from the subtraction of the two Nod positions and
extracted using the SSC program \emph{Spice}, or from the default
Nod1 - Nod2 difference spectra provided by the SSC. The error bars are calculated by combining the errors provided by the SSC with 2$\%$ of the photospheric flux at each wavelength.  A superflat was
created for groups of $\sim$15--20  stars in nearby IRS campaigns 
(Table~\ref{superflattable}), grouping stars by the date their data were taken.  Each superflat was derived by taking the ratio of the
SSC-spectra to Kurucz models \citep{kurucz92} appropriate for the effective temperature and metallicity of each star fitted to near-IR and
visible photometry as described in \citet{bryden06} and \citet{beichman06IRS}.  Stars with obvious excesses in
the IRS data or with excesses in the MIPS data (when available) were
excluded from the superflat. A few objects with problems in the IRS
spectra, e.g.,\ another star near the slit or obvious pointing
problems, were also rejected. To increase the sample size in making
superflats, we used IRS data from this sample and two closely related surveys 
that were taken at around the same time: the SIM/TPF sample \citep{beichman06SIM}, and the
FGK sample \citep{beichman06IRS}. Each module was normalized to the photospheric model using a single
constant whose value differed from unity by less than 25\% with a
dispersion of 8\%. The spectral data for each star in a group were
then divided by the group's superflat at each wavelength, thereby
eliminating any of the residual flat-field errors missed by the
standard \emph{Spitzer} pipeline reduction, including the ``droop'' at
$\sim$12 $\micron$ which was a significant source of error in some
of our brightest stars. As shown in Figure~\ref{fracex_noex}, this
process produces very uniform spectra, with the average fractional
excesses $[F_\nu($Observed$) - F_\nu($Photosphere$)] /F_\nu($Photosphere$)$ of all the stars used to make the superflats deviating from
zero by less than 0.5$\%$.

The defining characteristic of the dozens of debris disks we (and
others) have examined is an excess that first becomes detectable at
some minimum wavelength (typically longward of $\sim$20 $\mu$m, 
in the IRS LL1 or LL2 modules) and then deviates more and more from the
photosphere, rising to longer wavelengths. To look 
for weak excesses we calculated a multiplicative
calibration factor for each star and each IRS module using the first
10 data points in each module to ``pin'' the short wavelength end of
each module to the photospheric model. While the origin of these
residual gain errors is unknown (errors in the photospheric
extrapolation, stellar variability, or residual calibration errors
are all possible), the values of this calibration factor are small
and uniformly distributed around unity: $1.00\pm0.07$.

Three stars, HD 10360, HD 162004, and HD 185144, had calibration
factors significantly outside this range. Examination of 2MASS
images with the IRS slit superimposed showed that HD 10360 and HD
162004 had close companions that were in or close to the IRS slit
when data were taken. The AOR for HD 185144 was improperly aligned
with the slit, passing over the edge of the star rather than the
center, causing the flux to be improperly measured in the SL1
module. There was no evidence of an excess for any of these three
stars, albeit at a reduced level of precision ($<$5--10$\%$).

The technique of calibrating each module to the star's photosphere
produced smaller residuals and showed no significant deviation
from zero over the entire IRS wavelength range for the vast majority
of the sample. Defining, for convenience, two ``photometric bands''
useful for isolating either the silicate features (8.5--12 $\micron$)
or a long-wavelength excess (30--34 $\micron$), we see that the
dispersion in the deviation from a smooth photosphere was reduced
from $\sim$8$\%$ to 1$\%$ (8.5--12 $\micron$) and 2$\%$ (30--34
$\micron$) when examining non-excess stars.  We found no deviation between the stellar photosphere
and the IRS data for the majority of the sample, nor did we see any strong
evidence of silicate features in any of the stars (8.5--12
$\micron$).  We did, however, find clear evidence of excesses
longward of $\sim$15--25 $\micron$ for 16 stars and hints of a feature at $\sim$20 $\mu$m
for HD 10647 and HD 40136. 

In applying our technique we were very careful not to artificially
suppress any excess by our method of pinning the short wavelength
end of a module to the photospheric model. For example, for any
stars showing even a small excess in LL2 (14--21 $\mu$m), we adjusted the short end of
LL2 to fit the photosphere, and then adjusted the LL1
spectrum (21--34 $\mu$m) to fit the LL2 spectrum in their region of
overlap with a single gain term. If the SL1 spectrum (7--14 $\mu$m) showed
any hint of excess emission (this was only the case for one star: HD
219623), we tied LL1 $\rightarrow$ LL2 $\rightarrow$ SL1, and
anchored the short wavelength end of SL1 to the photosphere. In this
way, we proceeded from longer to shorter wavelengths ensuring that
no potential excess was lost.  

Splicing the modules together in this way does not necessarily produce results consistent with other methods of combining the modules.  HD 10647, which has the largest fractional excess of any of our sample stars, has its LL1 module spliced to the end of the LL2 module, which gives an excess of $96.4\pm2.8$ mJy in the 30--34 $\micron$ band.  \citet{chen06} found an excess of $114\pm2$ mJy in the same band for this star, implying that these error bars should be inflated when comparing excesses between surveys.  

We should note, however, that any excess from very hot dust, with
roughly a Rayleigh-Jeans spectrum at IRS wavelengths,  would be lost
in this procedure. This very hot dust has been invoked to account
for a spatially resolved excess at 2.2 $\micron$ observed by the Palomar Testbed Interferometer (PTI)
and Center for High Angular Resolution Astronomy (CHARA) interferometer \citep{ciardi01, absil06}.  Thus, we cannot rule out the existence of
material much hotter than 1000 K around any of these stars.  

Seven stars in the sample had an additional IRS measurement from
either the FGK or SIM/TPF samples (Table~\ref{pairtable}). 
For each of these stars we co-added the measured flux
at each wavelength, which reduced the noise and allowed us to remove
bad pixels. Three of these stars (HD~185144, HD~190406, and HD~222237) have no excess and this was confirmed by comparing the two
separate datasets.  Interestingly, HD~185144, which was tagged as a bad measurement because of low SL1 values due to improper slit alignment, also had low SL1 values in its
redundant measurement. 
Out of the remaining four stars, three (HD~115617, HD~158633,
and HD~199260) have excesses that were confirmed in the separate
datasets, and one star (HD~117043) has a weak excess after coadding.

%%%%%%%%%%%%%%%%%%%%%%%%%%%%%%%%%%%%
\subsection{SL2 and SL3 Analysis}

We examined the shortest wavelength data (SL2 and the ``bonus''
order, SL3)  using the same technique as described above. We
adjusted the SL3 data to fit SL2 and then tied the short wavelength
end of SL2 to the photospheric model. The dispersion ($1\sigma$)
around fits to the photospheric models is 1\% in a photometric
band defined between 6 and 6.5 $\mu$m and 2\% in a photometric
band defined between 7.5 and 8 $\mu$m. There was no evidence of any
excess shortward of 8 $\mu$m above the 3$\sigma$ level. 

\begin{figure}
\includegraphics[scale=0.65]{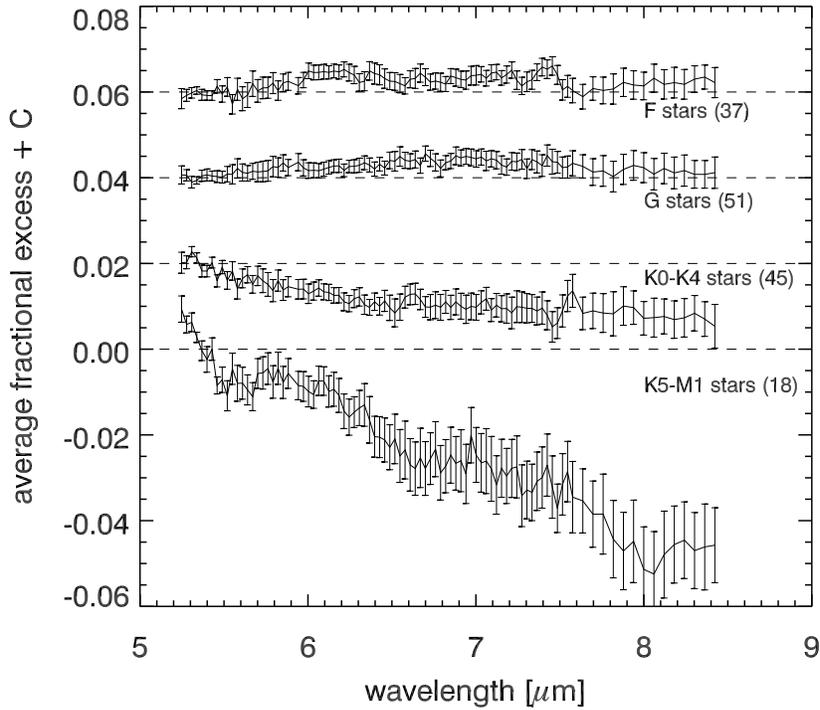}
\caption{Fractional excess after applying the superflat calibration
for 151 stars with valid SL2/SL3 data binned by spectral type. The
spectra have been pinned to the photospheric model at 5 $\micron$ so
deviations show up only at longer wavelengths. The small deviations
from zero  for the F and G stars demonstrate that the models are
very well behaved for these spectral types whereas deviations at the
3--5\% level are apparent at the longest wavelengths for the latest
spectral types. Similar deviation from a simple Rayleigh-Jean's
extrapolation is seen in the 24 $\micron$
photometry of late spectral types \citep{gautier07}.
\label{fracexSL2}}
\end{figure}

In performing the fitting we found that there was a systematic offset
between the Kurucz models and {\it Spitzer} spectra for stars later than
K5. Figure~\ref{fracexSL2} shows the fractional excess in the
SL2/SL3 wavelength band relative to the Kurucz models pinned to the
stellar emission at 5 $\micron$,  for four groups of spectral types:
F, G, K0--K4, and K5 and later. F and G stars reproduce the Kurucz
photospheres very clearly, while early K show small deviations
($\sim$1\%) and late K and early M stars show greater deviations
($>$3\%).  As reported by \citet{bertone04}, both of the commonly used stellar atmosphere models, Kurucz and NextGen \citep{hauschildt99a, hauschildt99b}, fail to accurately match later spectral type stars. 
From our analysis here and from previous investigations which used MIPS 24 $\mu$m observations of
nearby K and M stars 
\citep{beichman06SIM, gautier07} and found redder
$K_s-[24]$ colors than predicted by theory for both Kurucz and NextGen, it appears that this deviation between model and actual spectra is most severe closer to the near infrared. 
Examination of the
longer wavelength emission for these later type stars (IRS modules SL1 and LL1/2,
and MIPS data when available) revealed no evidence for longer
wavelength excess emission. Thus, we attribute this disagreement,
resulting in a 5--10\% apparent excess, as due to problems with the
photospheric models in the 2--10 $\mu$m portion of the spectrum, and
not as real excess due to dust emission.

\subsection{MIPS photometry} \label{MIPSphot}

While the focus of this paper is IRS spectra, for many of our sample
stars there is corresponding MIPS photometry at both 24 and 70 $\micron$.  
Most of this data has already been published,
for consistency we have re-reduced all of it with a uniform set of
analysis parameters.
Our analysis is similar to that previously described in
\citet{beichman05FGK}, \citet{bryden06}, and \citet{beichman06SIM}. 
At 24 $\micron$, images are created from the raw data
using software developed by the MIPS instrument team
\citep{Gordon05}, with image flats chosen as a function of 
scan mirror position to correct for dust spots 
and with individual frames normalized to
remove large scale gradients \citep{Engelbracht07}.
At 70 $\micron$, images are also processed with the 
MIPS instrument team pipeline which includes
corrections for time-dependent transients
\citep{Gordon07}. 
Aperture photometry is performed as in \citet{beichman05FGK}
with aperture radii of 15\farcs3 and 14\farcs8,
background annuli of 30\farcs6-43\farcs4 and 39\farcs4-78\farcs8,
and aperture corrections of 1.15 and 1.79 
at 24 and 70 $\micron$ respectively.
For three systems that are marginally resolved at 70 $\micron$  
(HD 10647, HD 38858, and HD 115617; see $\S$ \ref{othermodels2}), the small aperture fails to
capture all of the extended emission; for these three cases the MIPS 
fluxes listed in Table~\ref{MIPStable} are based on model fits to each
disk \citep{bryden09}. 
While our procedure has changed little since \citet{bryden06}
was published, note that improvements in the instrument calibration
since then have increased the overall 70 $\micron$ 
flux conversion by 4\%, from 15.8 to 16.5 mJy/arcsec$^2$/MIPS\_70\_unit
\citep[MIPS\_70\_unit is an internally defined standard based on the ratio of
the measured signal to that from the stimulator flash signal;][]{Gordon07}.
Overall, we find no qualitative disagreement between our results and 
those from earlier publications.

%%%%%%%%%%%%%%%%%%%%%%%%%%%%%%%%%%%%
\section{Results} \label{results}

\begin{figure}
\includegraphics[scale=0.7]{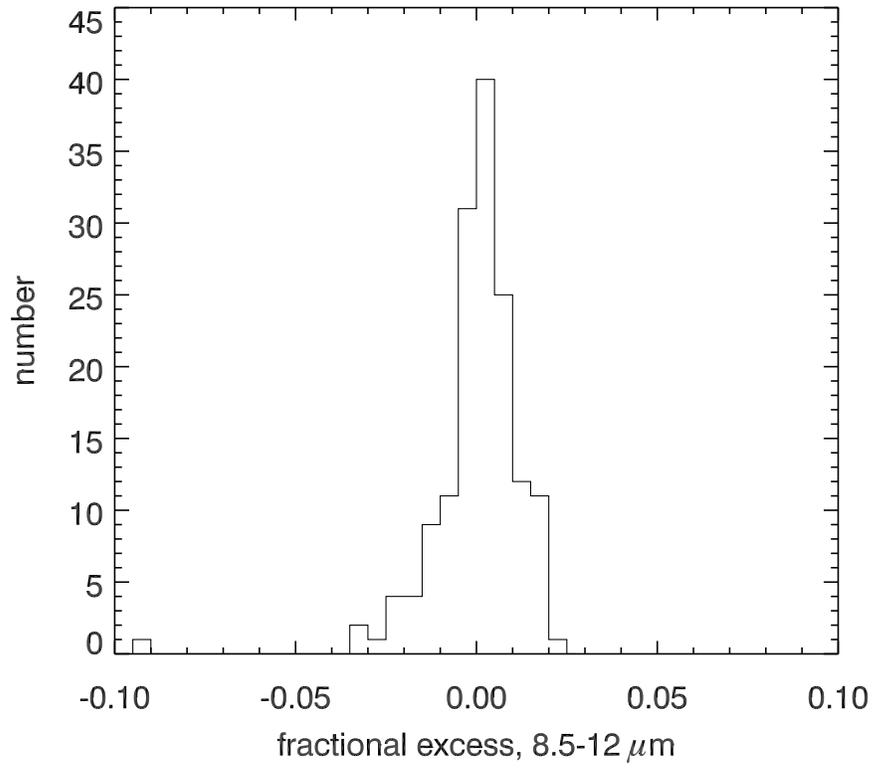}
\caption{Histogram showing the distribution of fractional excess
measured within a photometric band at 8.5--12 $\micron$.
\label{frac812hist}}
\end{figure}

\begin{figure}
\includegraphics[scale=0.7]{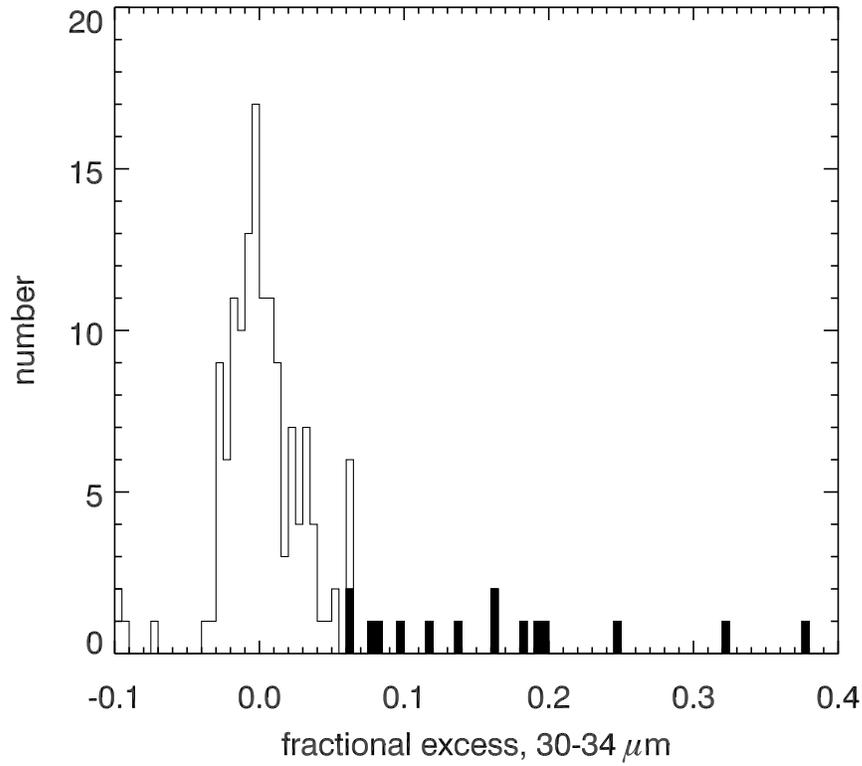}
\caption{Histogram showing the distribution of fractional excess
measured within a photometric band at 30--34 $\micron$. Filled bars
represent stars with statistically significant excesses. The point for HD 10647 is off-scale to the right with a fractional excess of 1.2. \label{frac3034hist}}
\end{figure}

After flattening and normalizing the IRS spectra as described above,
we estimate the fractional excess $[F_\nu($Observed$) -
F_\nu($Photosphere$)] /F_\nu($Photosphere$)$. We will continue to use
the two photometric bands previously defined to isolate either the
silicate features (8.5--12 $\micron$) or a long-wavelength excess
(30--34 $\micron$). Figures~\ref{frac812hist} and~\ref{frac3034hist}
show histograms of the fractional excess measured in these
photometric bands. In assessing the significance of an excess we
looked at the internal uncertainty in the flux density measurement
of a given star and the fractional excess relative to the $\sim$2\%
dispersion in the entire sample (Table~\ref{IRStable}).  The
amplitude of  the fractional excess relative to the entire
population is more important in assessing the reality of an excess
than the internal signal to noise ratio (S/N) in an individual
spectrum. There are a number of stars that appear to have a
significant excess when looking only at the internal uncertainties,
but which are not so impressive when compared to the dispersion in
the overall population. To assess the significance of any possible
excess we define $\chi$10 and $\chi$32 as $[F_\nu($Observed$) -
F_\nu($Photosphere$)] /$Noise for the two photometric bands, where Noise
is a combination of the dispersion in the fractional excess of the
individual spectrum and the population-averaged dispersion: 1$\%$
(8.5--12 $\micron$) and 2$\%$ (30--34 $\micron$). For a star to have an excess, we require
$\chi>3$ for an IRS-only detection or $\chi>2$ if the star also has
a MIPS 70 $\micron$ excess. Based on the data presented in
Table~\ref{IRStable} we can claim statistically significant 30--34
$\micron$ excesses for 16 stars (Table~\ref{MIPStable}). By this
same criterion, no stars in the sample have a significant 8.5--12
$\mu$m excess.

Complete IRS data for all 16 stars with excesses in these wavelengths are presented in the Appendix.  Figure~\ref{irsdataplot_noex} shows the IRS spectra for four representative stars
without excesses, while Figure~\ref{irsdataplot1} shows the IRS spectra
for all stars that do have significant excesses in the IRS wavelengths.   
The dotted lines in the right hand panels of Figures~\ref{irsdataplot_noex} and \ref{irsdataplot1} show an estimate of the 2$\sigma$ dispersion in the 
deviations from the photospheric models based on the entire sample; 
deviations between these lines should be regarded with skepticism.

%%%%%%%%%%%%%%%%%%%%%%%%%%%%%%%%%%%%
\subsection{Statistics of Detections} \label{statdetect}

We detected IRS excess emission toward 16 stars. These excesses
begin longward of $\sim$25 $\micron$ for 10 stars, and 
between $\sim$15--25 $\micron$ for the other 6 stars.  Two of the excess detections are
of borderline significance and are included because of the additional information of a MIPS 70 $\micron$ excess (see $\S$ \ref{mipsdisc}): HD~110897 and HD~117043, both with $\chi$32 = 2.8. 
Out of the sample of 152 stars, these 16 stars correspond to a 30--34
$\micron$ excess detection rate of 
$10.5\%\pm2.6\%$, which is consistent with the fraction of stars
with excesses found in a previous IRS survey: $12\%\pm5\%$
\citep{beichman06IRS}.  We must however, correct these statistics
for the sources that were not observed as part of this sample
because they were claimed as part of other, earlier \emph{Spitzer} programs.
Comparing our initial selection of
sources meeting our astrophysical criteria with early Guaranteed
Time or Legacy programs yields 51 additional stars,
which we list in Table~\ref{rejecttable}.  With this
correction, the success rate for long-wavelength IRS excesses is not
$16/152=10.5\%\pm2.6\%$, but $24/203=11.8\%\pm2.4\%$, essentially the
same as found in our earlier determination \citep{beichman06IRS}, 
but with much lower uncertainty. 

HD~72905, from the FGK survey \citep{beichman06IRS}, presents an interesting example of the challenges in identifying a weak  infrared excess, particularly around  8-14 $\mu$m where the stellar photosphere is bright. Using IRAC data from the FEPS program \citep{carpenter08} we use our standard technique to fit  Hipparcos visible photometry, partially saturated 2MASS observations at JHK$_s$,  and IRAC 3.6 and 4.8 $\mu$m data to a Kurucz model for a 6,000~K G0V star with [Fe/H] = -0.08. The resultant fit has a reduced $\chi^2$ of 0.97.  Pinning the SL2 data to a Kurucz photosphere using the 20  shortest wavelength  SL2 points requires a $\sim$2\% adjustment to the SSC pipeline data and reveals {\it no fractional excess}  from 5-8 $\mu$m greater  than 2\%.  A similar conclusion applies if we fit a solar photosphere \citep{rieke08} to the IRAC 3.6 and 4.8 $\mu$m data. Extending  the Kurucz photosphere  to longer wavelengths yields  a marginal excess of about 5\% at IRAC 7.8 $\mu$m that carries through to IRS SL1 and MIPS 24 \micron.  The fractional excess has a significance at the $\sim$2$\sigma$ level relative to the $\sim$2\% uncertainties in the photospheric models. However, changing photospheric models makes the excess all but vanish. Fitting the \citet{rieke08} solar photosphere instead of the Kurucz model reduces the level of excess to 2\% or less out to 25 $\mu$m (including MIPS 24). We conclude that we cannot claim any  statistically significant excess at $<$25 $\mu$m. At longer wavelengths, the difference between photospheric models becomes less important and the existence of a weak excess starting at $\lambda>25\mu$m becomes evident.

None of our sample stars showed excesses in the short wavelength 8.5--12
$\micron$ portion of the spectrum, giving a fractional incidence of
$<$0.7\% for these mature stars. 
Adding in stars with previous
{\it Spitzer} observations,
we find an overall excess detection rate of 2 stars (HD~69830 and HD~109085) out
of 203 = $1.0\%\pm0.7\%$ for the 8.5--12 $\micron$ band. This confirms the
rarity of detectable short-wavelength excesses compared with ones at longer
wavelengths, as seen in \citet{beichman06IRS} and noted in earlier studies using the 
Infrared Astronomical Satellite (IRAS) and the Infrared Space Observatory (ISO). 

The FEPS survey \citep{carpenter08, hillenbrand08} used {\it Spitzer} to observe nearby sun-like stars with ages between 3 Myr and 3 Gyr.  Using our criteria of $>$3$\sigma$ above the photosphere (or $>$2$\sigma$ with a known 70 $\micron$ excess), \citet{hillenbrand08} find excesses in the 30--34 $\micron$ band for 22 out of the sample of 328 stars, although not all stars in the sample have reported IRS spectra.  \citet{carpenter08} measure excesses using colors rather than comparison with Kurucz models, and find 71 out of 314 stars ($22.6\%\pm2.7\%$) with excesses in the long wavelength IRS band, and 2 out of 314 ($0.6\%\pm0.5\%$) in the short wavelength IRS band.  As these stars are on average younger than the stars in our sample, it is not surprising that there is a higher incidence of IRS-detected excesses.  

Five stars in the sample were known to have planets as of May 2009: HD~4308 \citep{udry06}, HD~10647 \citep{mayor03}, HD~40307 \citep{mayor09}, HD~154345 \citep{wright08}, and HD~164922 \citep{butler06}.  Of these five stars, only  HD~10647 shows an excess at both 70 $\micron$ and IRS wavelengths. The other four planet-bearing systems have no detected excesses.

%%%%%%%%%%%%%%%%%%%%%%%%%%%%%%%%%%%%
\subsection{Discussion of MIPS Results} \label{mipsdisc}

We have MIPS data from other programs for about half (78) of the sample stars, as noted in Table~\ref{IRStable}. Table~\ref{MIPStable} lists all of the stars in our sample with IRS and/or MIPS 70 $\micron$ excesses. Of the 16 stars with IRS
excesses, 14 have excesses in both the 30--34 $\micron$ IRS band and the MIPS photometry at 70 $\micron$; only
HD~154577 has an IRS excess with no detectable MIPS excess (HD~190470 is in a particularly noisy field close to the galactic plane, so the error bars on the 70 $\micron$ flux are so large that nothing can be said about whether or not there is an excess). 
Including the stars with previous IRS observations (Table~\ref{rejecttable}) gives 22 stars with IRS excesses, 20 of which also have strong or weak MIPS 70 $\micron$ excesses.  HD~110897 was not originally considered to have an IRS excess because of a 
marginal $\chi$32 value (2.8), but this can be considered a weak excess because of the additional information of a strong MIPS 70 $\micron$ excess. HD~117043 has a
marginally significant IRS excess ($\chi$32 = 2.8) and a
marginally significant MIPS 70 $\micron$ excess ($\chi$70
= 1.9), but because $\chi$32 is close to 3, this star was
also included as a weak excess detection at both wavelengths. 

Out of the 73 stars with both MIPS 70 $\micron$ and IRS data, only three stars (HD~90089, HD~132254, and HD~160032) have excess MIPS 70 $\micron$ emission with no significant IRS excess.  \citet{hillenbrand08} finds a similar trend, with $>$80$\%$ of their stars with MIPS 70 $\micron$ excesses also possessing IRS 33 $\micron$ excesses, and no reported stars possessing an IRS excess with no corresponding MIPS 70 $\micron$ excess.  This implies that there may be a lower limit to debris disk temperatures, with a corresponding upper limit on disk sizes.  Kuiper belt analogs appear to happen preferentially in regions with temperatures around 50 K, and not at lower temperatures.   

All of the stars with MIPS 70 $\micron$ data also have MIPS 24
$\micron$ measurements. Only two stars have greater than 2$\sigma$ 24
$\micron$ fractional excesses: HD~10647 has a 24 $\micron$ excess that agrees with its large IRS excess. HD~38392 has a large apparent
24 $\micron$ excess with no IRS excess. However, examination of the 2MASS and
MIPS 24 $\micron$ image for this star shows a bright companion
star, HD 38393 ($K_s \sim$ 2.5 mag), about 1\farcm5  away.
Although the IRS slit does not cross the companion star, and therefore should not effect the spectrum, the uncertainty in the 24
$\micron$ photometry is inflated by the companion. Further, this
star appears to be variable at the 5\% level in a number of visible
compilations \citep[Hipparcos time series photometry and][]{nitschelm00}. An alternate explanation is the presence of an M dwarf companion.  Such a companion could produce a 20$\%$ excess at 24 $\micron$, and would be too faint to notice if the system's spectral type was measured using optical observations.  Follow-up imaging using adaptive optics would be needed to test this hypothesis.  Reinforcing the peculiarity of the MIPS 24 $\micron$ datapoint, the MIPS data do not show any 70 $\micron$  excess for HD 38392.

%%%%%%%%%%%%%%%%%%%%%%%%%%%%%%%%%%%%
\subsection{Limits on the Fractional Disk Luminosity} \label{lumlimit}

A useful metric for the limits on dust surrounding these stars is
$\ld$, which is related to the fractional flux limit of an
excess relative to the Rayleigh Jeans tail of the star's photosphere
~\citep{bryden06, beichman06IRS}:

\begin{equation}
{ {L_{\rm dust}} \over {L_*} } = { {F_{\rm dust}} \over {F_*} } { {
e^{x_d}-1} \over {x_d} } \left ( {{T_d} \over {T_*}}\right ) ^3
\label{dusteq}
\end{equation}

\noindent where $F_{\rm dust}=F_\nu($Observed$)-F_\nu($Photosphere$)$. At the peak of the blackbody curve $x_d\equiv h\nu/kT_d$ has a
constant value of 3.91, corresponding to $T_d=367$ K at 10
$\micron$. At this wavelength $\ld = 3.5 \times 10^{-3} \, ( T_* /5600 \, {\rm K} ) ^{-3} 
\fd$. At 30--34 $\micron$ the
corresponding equation is  $\ld = 1.3\times10^{-4} \, ( T_* /5600 \, {\rm K} ) ^{-3} 
\fd$, assuming $T_d=115$ K. (For comparison, the typical dust temperatures traced by the MIPS 24 and 70 $\micron$ data are 154 and 53 K, respectively.) In
Table~\ref{IRStable} and Figure~\ref{LdLstarhist} we evaluate
$\ld$ for each star using the appropriate effective temperature (listed in Table~\ref{basictable}), luminosity 
(from our stellar photosphere models),
and its
measured fractional excess in each band, $F_{\rm dust}/F_*$, or, in the case of an
upper limit, $3\sigma_{pop}$ where $\sigma_{pop}$ is the dispersion
in fractional excess averaged over the whole sample (0.010 at 8.5--12 $\micron$; 0.028 at 30--34 $\micron$). This definition
of $\ld$ assumes that the emitting material is all at the
location where the peak of the $T_d$ blackbody matches the
wavelength of observation  such that for stars with excesses the given value of $\ld$ is
actually a minimum. More dust emission, and higher values of
$\ld$, would be required for material located substantially
interior or exterior to this point.

The 3$\sigma$ limits on $\ld$ at 8.5--12 $\micron$ and 30--34
$\micron$ have 2$\sigma$ clipped average values of $\ld=11\pm
4\times 10^{-5}$ and $1.31\pm 0.49 \times 10^{-5}$ respectively 
(Table~\ref{IRStable}). In comparison with our solar system, which has
$\ld \sim$ 10$^{-7}$ \citep{bp93,dermott02}, the IRS results set
limits (3$\sigma$) on warm (360 K) dust peaking at 10 $\micron$ of
$\sim$1,000 times the level of dust emission in our solar system. For
cooler dust ($\sim$115 K) peaking at 30--34 $\micron$, the 3$\sigma$
limit corresponds to $\sim$100  times the nominal $\ld$ of our
zodiacal cloud.  For objects with excesses in the IRS bands, we determine
$\ld$ explicitly by integrating over the data between 10--34
$\micron$ and using the models discussed below ($\S$ \ref{simplemodel}) to extrapolate out to and beyond
the MIPS 70 $\micron$ datapoint.

%%%%%%%%%%%%%%%%%%%%%%%%%%%%%%%%%%%%
\subsection{Comparing IRS and MIPS Statistics}

Figure \ref{fluxrates} summarizes the rates of IR excess detection in IRS spectral surveys and compares them with MIPS photometric results.
For two wavelengths in each instrument, the distribution of detection
rates is shown as a function of the fractional dust flux
($\fd$).   Note that $\fd$ can be easily
translated to a fractional disk luminosity using Equation~\ref{dusteq}.
It is clear from this figure that the dominant dust around solar-type
stars tends to be colder than is optimal for detection at IRS wavelengths and
generally exhibits higher $\fd$ at longer wavelengths.
Nevertheless, because we can detect excesses down to much smaller levels
of $\fd$ within the IRS spectra, the overall detection rate
of IR excess for IRS at 32~$\micron$ is similar to that for MIPS at 70~$\micron$.
By comparison, IRS at 10~$\micron$ and MIPS at 24~$\micron$ have relatively few
detections, but are both consistent with the overall trend from the
other wavelengths. 
While it is difficult to extrapolate these distributions down to
fainter values, the curves
can be fit by log-normal distributions with median values of
$\fd \sim$ 0.06 at 70~$\micron$ and
$\fd \sim$ 0.003 at 32~$\micron$.
%***asdf
These fractional fluxes correspond to $\ld \sim 5 \times 10^{-7}$
for a solar temperature star, consistent with estimates for our Kuiper
Belt's emission \citep{stern96}.

While the individual spectra provide the best measure of the range of
dust temperatures in each system ($\S$ \ref{spectrachar}), Figure~\ref{fluxrates}
provides a sense of the generalized disk characteristics.
The separation between the 32 $\micron$ and 70 $\micron$ distributions
in Figure \ref{fluxrates}, for example, 
can be translated to a representative dust temperature of $\sim$65 K.
In reality a range of temperatures are present and,
as is found in $\S$ \ref{simplemodel}, the dust in each system is
often not well fit by a single emission temperature, but rather 
by a distribution.
This is also apparent from the overall statistics,
as evidenced by the inability of a single blackbody to fit the trends
seen in Figure~\ref{fluxrates}; the separation between the 24 and 70 $\micron$
curves is consistent with 75 K dust, while the separation between the 10
and 70 $\micron$ curves corresponds to dust temperatures $>$100 K.
A similar trend is found by \citet{hillenbrand08},
who find that $>$1/3 of their surveyed debris disks have evidence for
multiple dust temperatures based on their colors at MIPS and IRS wavelengths.

%%%%%%%%%%%%%%%%%%%%%%%%%%%%%%%%%%%%
\section{Discussion} \label{discussion}

%%%%%%%%%%%%%%%%%%%%%%%%%%%%%%%%%%%%
\subsection{Characteristics of the Spectra} \label{spectrachar}

The excesses found in this survey are in most cases weak and relatively featureless beyond a simple rise to longer wavelengths. A few objects are exceptional: HD~10647 stands out for having a very strong excess, $\ld=10^{-3.9}$ rising up to 70 $\micron$ and continuing out to 160 $\micron$ \citep{tanner08}; this source also appears to be extended at 70 $\micron$ \citep{bryden09}.  HD~40136 and possibly HD~10647 show a bump around 20 $\micron$ which might be attributable to small grains.  In addition to HD~10647, HD~38858 and HD~115617 also both show evidence for extended MIPS emission \citep{bryden09}.

%%%%%%%%%%%%%%%%%%%%%%%%%%%%%%%%%%%%
\subsection{Models for the Dust Excesses} 

The IRS and MIPS excesses detected toward some of the 152 stars discussed here can be used to characterize the properties and spatial location of the emitting material. Unfortunately, even the simplest characterization cannot be unique given the wide variety of grain sizes and compositions as well as possible locations for these different species. The complexity of debris disks is evident as one attempts to model the most prominent debris disks for which high-quality IRS spectra and fully resolved maps are available \citep[e.g. Vega;][]{su05}. In this section we first apply a simple, single-component model that fits the majority of sources; we assume uniform, large-grained ($\sim$10 $\micron$) dust is located in an annulus centered on the star. We then examine somewhat more sophisticated models for disks where additional complexity seems warranted.

\subsubsection{Simple Dust Models} \label{simplemodel}

As a first step in analyzing these data we fitted the IRS spectra
and MIPS 70 $\micron$ photometry using a simple model
of optically thin dust located within a single dust annulus centered
around the star.  As described in \citet{beichman06IRS} we calculated the power-law
temperature profiles, $T(r)=T_0 (L/L_\odot)^\alpha (r/r_0)^\beta$,
for grains in radiative equilibrium with the central star. 
We use dust emissivities for 10 $\micron$ silicate grains
\citep{draine84, weingartner01}, the minimal size suggested 
by the lack of significant features in most of the spectra.
For the 10 $\micron$ silicate grains we obtained the following numerical relationship: 
$T(r)=255 \, K (L/L_\odot)^{0.26} (r/AU)^{-0.49}$. 
These calculated coefficients
and power-law constants closely follow analytical results
\citep{bp93}. We then calculated the dust excess by integrating over
the surface brightness of a disk between $R_1$ and $R_2$, with
$F_\nu(\lambda)={ {2\pi} \over {D^2} } \int \tau_0(\lambda)
(r/r_0)^{-p} B_\nu(T(r)) r dr $. The disk surface density 
distribution expected for grains dominated by
Poynting-Robertson drag is roughly uniform with radius, i.e.,\ $p=0$
\citep{burns79, buitrago85, backman04}. We examined a number of
other cases with $0<p<1$ that would reflect different dust dynamics,
but did not find results that were substantially different from
those for $p=0$.

We fitted the excess emission from a single annulus to 83 data
points longward of 21~$\micron$ (just longward of the last point
used for flux normalization of the LL1 IRS module) for 10 stars,
and to 160 data points longward of 14~$\micron$ (just longward of
the last point used for the flux normalization of the LL2 IRS
module) for 9 stars, depending on whether  there was any hint of
an excess shortward of 21 $\micron$. We  included the MIPS 70 $\micron$
data, which was available for all 19 of the stars modeled.  By varying $\tau_0$, $R_1$ and $R_2$ we
were able to minimize the reduced $\chi^2$ to values between 0.6--1.2, 
except for  HD~10647, which has a fit with a reduced $\chi^2$ of 5.7, indicating a simple 10 $\micron$ dust grain model does not satisfactorily fit the infrared excess observed for this star (see $\S\S$ \ref{othermodels1} and \ref{othermodels2}). 
Results of the model fitting are shown in Figure~\ref{allexplotall} and Table~\ref{dusttable} and are discussed
below.

Mass estimates are notoriously tricky to derive given uncertainties
in grain sizes. Assuming a silicate grain density of 3.3 g
cm$^{-3}$, we calculate dust masses of 4~$\times~10^{-7}$--2.4~$\times~10^{-3}
M_\oplus$. Extrapolating this estimate using
the $-3.5$ index power-law appropriate for a distribution of sizes
from a collisional cascade \citep{dohnanyi69} up to a maximum size of 10 km yields total mass estimates as shown in Table~\ref{dusttable}.  Submillimeter
observations of all these sources would further constrain the dust
size and distribution and thus the total mass of the emitting
material.  

$\ld$ values were obained by integrating the excess over frequency, including a power law interpolation between 35 and 70 $\micron$ (if available). We used a simple blackbody curve to extrapolate beyond 70 $\micron$, based on the middle of the temperature range found by the model and quoted in Table~\ref{dusttable}.  For stars with a 70 $\micron$ excess only, we used Equation~\ref{dusteq} at 70~$\micron$.

The models match the spectra quite well (Figure~\ref{allexplotall}), yielding dust temperatures between
$\sim$50--450 K (Figure~\ref{t1t2}) and dust locations between
$\sim$1--35 AU (Figure~\ref{r1r2}). The majority of disks are
located between 10 and 30 AU from their stars with several ($\sim$7/19) showing a single temperature fit perhaps indicative of a
ring-like structure that may be found with higher resolution data.
 Since the IRS data place only a limit on the emission shortward of
34 $\micron$ for HD 90089, HD 132254, and HD 160032, the single
grain model can be used to show that the inner edge of the disk seen
at 70 $\micron$ must start beyond $\sim$15 AU corresponding to
material cooler than $\sim$70K. The mean value of the disk sizes shown in Figure~\ref{r1r2} is
14 $\pm$ 6 AU. 

It is important to note, however, that these disk sizes are crucially
dependent on the assumed grain size and that, as discussed below,
smaller grains could dramatically increase the distance at which the
emitting grains are actually located \citep[e.g.,][]{bryden09}.

%%%%%%%%%%%%%%%%%%%%%%%%%%%%%%%%%%%%
\subsubsection{More Complex Dust Models} \label{othermodels1}

While models using single population of 10 $\mu$m dust grains reproduce  the weak, featureless spectra of most of our stars with excesses, we  tried to model some of the excesses using a more realistic mixture of grains of different sizes and material compositions. The compositional model applied to HD~69830 \citep{lisse07} and HD~113766 \citep{lisse08} utilizes a combination of water ice, amorphous and crystalline olivine, and amorphous and crystalline pyroxene.  The mix used here contains roughly 50:50 rocky dust and water ice, similar to the abundances seen in the small icy bodies of the Kuiper Belt.  But of the 152 program stars, only two (HD~10647 and HD~40136) show hints for spectral features around 20 $\mu$m and  neither of these stars has a statistically significant IRS excess shortward of 18 $\mu$m, severely hampering the fitting due to the lack of emission features available to constrain the models. The remainder of the stars did not offer enough statistically significant data to merit more sophisticated modeling than the simple characterization described in \S\ref{simplemodel}. 

Applying the compositional model to HD~40136, we were able to derive good, although relatively unconstrained fits to the 18--35~$\mu$m IRS data, finding evidence for crystalline olivine (50:50 Fe/Mg rich), crystalline pyroxene, FeS and some water ice, with a reduced $\chi^2$~$\sim$~0.8 (Figure~\ref{caseyfigs}).  Removing silicates worsened the fit to a reduced $\chi^2$~$\sim$~1.6, mostly due to a failure to fit the data around the 18--20~$\micron$ silicate feature.  However, the signal to noise ratio is poor shortward of 20~$\micron$ due to the bright stellar photosphere, making these identifications preliminary. 

The model for HD~10647 yields a similar mix of ices and silicates, with a reduced $\chi^2$~$\sim$~0.8 (Figure~\ref{caseyfigs}).  Removing silicates from the fit gives a significantly worse reduced $\chi^2$~$\sim$~56, strongly supporting the inclusion of silicates in the model spectrum.  This model is discussed further in \citet{tanner08}.  

Because of the low signal to noise ratio shortward of 18 $\micron$ for both of these stars, identification of minerals will have to await future observations.  The {\it Herschel Space Observatory} could prove especially useful to check for the evidence of a water ice feature near 62~$\micron$ (Figure~\ref{caseyfigs}).

%%%%%%%%%%%%%%%%%%%%%%%%%%%%%%%%%%%%
\subsubsection{Stars with Observed Extended Emission} \label{othermodels2}

HD~10647, HD~38858, and HD~115617 are all marginally extended in their MIPS 70 $\micron$ images.  When dust rings are fit to this extended emission, \citet{bryden09} find much larger radii ($\sim$100 AU) than indicated by our models.  This can be explained by either different grain emissivities, or by two populations of dust grains: larger, 10 $\micron$ dust grains in a closer annulus ($\sim$10--30 AU), and smaller dust grains at larger radii ($\sim$100 AU)

HD~10647 and HD~115617 are also detected in MIPS 160 $\micron$ images \citep{tanner08}, further supporting the hypothesis of two dust populations.  At 160 $\micron$, emission from the stellar photosphere is negligible, so the detected emission is attributed to cold ($\sim$30 K) dust at large distances ($\sim$100 AU) from the star, much farther out than our model based on the warm dust predicts.    

The case of the planet-bearing star HD~10647 is particularly interesting since not only is it detected at 160 $\mu$m and resolved at 70 $\mu$m, its disk is also resolved in coronagraphic images from the Advanced Camera for Surveys on the \emph{Hubble Space Telescope} \citep{stapelfeldt07}. 
Its very high $\ld$ and young age make this the most likely star in our sample to have small grains due to a recent collisional event.  A compositional model (as used in $\S$~\ref{othermodels1}) incorporating an additional population of very cold, small grains composed primarily
of water ice fits the combined data sets very well. The data and the
relevant model are discussed in depth in \citet{tanner08}.

%%%%%%%%%%%%%%%%%%%%%%%%%%%%%%%%%%%%
\subsection{Characteristics of the Dust} \label{dustchar}

Only 3 stars have convincing evidence for warm dust: HD~40136, HD~190470, and HD~219623. One star, HD~117043, has hints of warm excess but is too weak at both IRS and MIPS for further consideration. HD~40136 and HD~219623 have MIPS excesses as well, while HD~190470 has significant cirrus contamination so the MIPS limit is poor.  The simple model ($\S$ \ref{simplemodel}) for HD~40136 and HD~219623 show material extending to within 1 AU (Figure~\ref{r1r2}), suggestive of disks with active reprocessing of material, given the short grain lifetimes at these small orbital radii \citep{wyatt08}.

All of the remaining stars with excesses have their emitting material located in regions analogous to the Kuiper Belt in our solar system, typically beyond 10~AU, out to a maximum value of 30~AU (Figure~\ref{r1r2}). In the two cases where the signal-to-noise ratio is (barely) adequate for mineralogical analysis, HD~10647 \citep[$\S$~\ref{othermodels1} and][]{tanner08} and HD~40136 ($\S$~\ref{othermodels1}), the suggestion of significant amounts of water ice is intriguing and is to be expected for regions that lie well beyond the snow line, where volatiles are predicted to be abundant \citep{pollack96}. Figure~\ref{r1r2} shows the location of the snow line for 1~Myr old stars \citep[using stellar models from][]{siess00}, an age when stellar luminosity and the volatile content of the outer disk should be stabilizing. The majority of our sample 
%(1 IRS only, 11 IRS/MIPS excesses or 3 MIPS only excess CHECK) 
have material located at or well beyond the snow line.

The total quantity of material responsible for the observed excesses is poorly constrained by our data, because of uncertainties in grain properties and in the extrapolation up to maximum particle size. Table~\ref{dusttable} lists total masses extrapolating a population of bodies with 3.3~g~cm$^{-3}$ and a $N(a) \propto a^{-3.5}$ size distribution up to 10~km. The median value for 13 stars with strong IRS and MIPS excesses is 0.34~$\MEarth$. The average and standard deviation, 1.0~$\pm$~2.1 $\MEarth$, are dominated by a few outliers with more massive disks: HD~1461, HD~38858, and HD~45184 all around 1--2 $\MEarth$, and especially HD~10647 with an extrapolated total mass of 7.7~$\MEarth$. These mass estimates can be compared to various estimates for the mass of our own Kuiper belt or to models of the primitive solar nebula. In the Nice model of the protosolar nebula, for example, the outer disk is predicted to contain roughly 10-150~$\MEarth$ of material in bodies with densities of 1~g~cm$^{-3}$ in sizes up to 300~km \citep{alessandro09}. It is difficult to compare these values directly to ours, since we assumed a smaller maximum size, 10~km, vs.\ 300~km for \citet{alessandro09}.  However, this is offset by our different density assumptions: we assumed 3.3~g~cm$^{-3}$, while \citet{alessandro09} used 1~g~cm$^{-3}$. More importantly, our 18--70~$\micron$ data are probably missing significant emission from other populations of dust: more distant and/or larger grains would emit at longer wavelengths. On the other hand, the order of magnitude agreement between our measurements and nominal solar system values is encouraging.

It should be noted however, that the stars with strong excesses are in the minority of our sample ($<$12\%) and that the vast majority of the mature stars in our study (and other {\it Spitzer} studies) have Kuiper Belt disk masses less than 0.1~$\MEarth$. This relatively strict upper limit must eventually be reconciled with the presence or absence of gas giant or icy giants in the outer reaches of planetary systems.

%%%%%%%%%%%%%%%%%%%%%%%%%%%%%%%%%%%%
\section{Conclusion} \label{conclusion}

We have used the IRS instrument on the \emph{Spitzer Space
Telescope} to look for excesses around nearby,
solar-type stars. We find that none of our 152 sample stars have
significant excesses in the 8.5--12 $\micron$ portion of the
spectrum, while 16 have excesses beginning at $\sim$15--25 $\micron$
and rising to longer wavelengths. Including stars that meet our sample criteria and were previously observed
with the IRS instrument, we find that $11.8\%\pm2.4\%$ of nearby,
solar-type stars have excesses at 30--34 $\micron$, while only
$1.0\%\pm0.7\%$ have excesses at 8.5--12 $\micron$. The rarity of
short wavelength excesses is consistent with models~\citep{wyatt07};
for ages older than 1 Gyr, disks should fall below our sensitivity
threshold. Bright emission such as that seen toward HD 69830 must be
intrinsically rare, have a duty cycle less than 1$\%$ of the
typical 2 Gyr age of these stars, or an occurrence less than once
per 20 million years. This could mean the habitable zones of nearby solar-type stars have a very low incidence of massive collisions, providing opportunity for stable, catastrophe-free terrestrial planets. 

\acknowledgments {
This publication makes use of data products from the Two-Micron All
Sky Survey (2MASS), as well as from IPAC/IRSKY/IBIS, SIMBAD, VizieR, the ROE Debris Disks Database website, and the Extrasolar Planets Encyclopaedia website.
The {\it Spitzer Space Telescope} is operated by the Jet Propulsion
Laboratory, California Institute of Technology, under NASA contract 1407. 
Development of MIPS was funded by NASA through the Jet Propulsion
Laboratory, subcontract 960785.  Some of the research described in
this publication was carried out at the Jet Propulsion Laboratory,
California Institute of Technology, under a contract with the National
Aeronautics and Space Administration. S.\ M.\ L.\ would like to thank Seth Redfield and Roy Kilgard for very helpful comments and advice regarding this paper.}

%%%%%%%%%%%%%%%%%%%%%%%%%%%%%%%%%%%%

\clearpage
\begin{deluxetable}{l|ll|ccccc|cccl|ccl}																																	
\tablewidth{0pt} \tabletypesize{\scriptsize}  \rotate \setlength{\tabcolsep}{0.03in} \tablecaption{Basic Data\label{basictable}}																																	
\tablehead{																																	
		&		&		&		&		&		&		&				&	\multicolumn{4}{c}{Age} 								&		&		&		\\\cline{9-12}
Star		&	HIP	&	GJ	&	Spectral	&	Temp.	&	Distance	&	V	&	K			&	W/V/Avg$^{a}$		&	Min	&	Max	&		&	\multicolumn{3}{c}{[Fe/H]}					\\\cline{13-15}
		&		&		&	Type	&	(K)	&	(pc)	&	(mag)	&	(mag)			&	(Gyr)		&	(Gyr)	&	(Gyr)	&	References	&	Average	&	$\sigma$	&	References	}
\startdata																																	
HD 870		&	1031	&	2001	&	K0V	&	5273	&	20.3	&	7.22	&	5.38			&	2.0		&	\nodata	&	\nodata	&	RP	&	-0.22	&	0.06	&	N,RP	\\
HD 1461		&	1499	&	16	&	G0V	&	5948	&	23.4	&	6.47	&	4.90			&	6.3		&	4.2	&	9.6	&	W,V,I,N	&	0.3	&	0.1	&	B,CS,Ce,I,M,N,P,T,Th	\\
HD 1581		&	1599	&	17	&	F9V	&	6017	&	8.6	&	4.23	&	2.78$^{b}$		&	5.3		&	4.1	&	8.2	&	V,I,N,RP	&	-0.2	&	0.1	&	CS,E,I,La,M,N,RP,T,Th	\\
HD 3765		&	3206	&	28	&	K2V	&	5047	&	17.3	&	7.36	&	5.16			&	7.0		&	4.8	&	7.0	&	V,RP	&	0.0	&	0.1	&	B,CS,I,M,N,P,RP,T	\\
HD 4308$^{c}$	&	3497	&	32	&	G5V	&	5678	&	21.9	&	6.55	&	4.95			&	9.5		&	8.6	&	9.6	&	V,I,N,RP	&	-0.3	&	0.1	&	B,CS,I,M,N,RP,T	\\
HD 4813		&	3909	&	37	&	F7IV-V	&	6226	&	15.5	&	5.17	&	4.02$^{d}$	&	2.8		&	1.4	&	3.8	&	C,I,La,L,N	&	-0.14	&	0.08	&	B,CS,C,I,La,L,M,N,P,T,Th	\\
HD 5133		&	4148	&	42	&	K3V	&	4925	&	14.1	&	7.15	&	4.89			&	6.0		&	5.4	&	6.0	&	V,I	&	-0.13	&	0.07	&	CS,I,M,N,T	\\
HD 7439		&	5799	&	54.2 A	&	F5V	&	6445	&	24.4	&	5.14	&	4.06$^{d}$	&	3.0		&	2.4	&	4.2	&	C,L,M,N	&	-0.31	&	0.06	&	B,CS,C,L,M,N,T,Th	\\
\enddata																																	
\tablecomments{Spectral Types from SIMBAD.  Visual magnitudes are as quoted in SIMBAD, typically from the Hipparcos satellite, and K magnitues are from 2MASS unless otherwise noted.}																																
\tablenotetext{a}{Age from \citet{wright04} or \citet{valenti05} paper if available, otherwise an average of literature values.}																																
\tablenotetext{b}{Star has J, H and/or K values from \citet{johnson01} or other literature.}																																
\tablenotetext{c}{Star has at least one known radial velocity planet.}																																
\tablenotetext{d}{Star has one or more bad 2MASS values (error $>$ 20$\%$), slightly reducing the accuracy of its photometric model.}																																
\tablenotetext{e}{After considering several factors, our age for HD 10647 comes from \citet{chen06} (see note in $\S\ref{sample}$). Minimum and maximum ages for this star are from all literature sources.}																																
\tablerefs{B: \citet{borkova05}; Ba: \citet{barry88}; Bo: \citet{borges95}; C: \citet{chen01}; Ch: \citet{chen06}; CS: \citet{cayrel97, cayrel01}; Ce: \citet{cenarro01}; D: \citet{ducati02}; E: \citet{eggen98}; F: \citet{feltzing01}; I: \citet{ibukiyama02}; K: \citet{kidger03}; L: \citet{lambert04}; La: \citet{lachaume99}; Le: \citet{lebreton99}; M: \citet{marsakov88, marsakov95}; Ms: \citet{mashonkina01}; Ml: \citet{morel78}; N: \citet{nordstrom04}; P: \citet{perrin77}; R: \citet{randich99}; RP: \citet{rocha98}; S: \citet{sylvester00}; T: \citet{taylor03}; Th: \citet{thevenin98}; V: \citet{valenti05}; W: \citet{wright04}}																																
\end{deluxetable}

\clearpage
\begin{deluxetable}{ccccc}
\tablewidth{0pt} \tabletypesize{\scriptsize}
\tablecaption{Superflats\label{superflattable}} \tablehead{
Superflat   &   IRS &   Programs    &   $\#$ stars using    &   $\#$ stars used     \\
    &   campaigns   &   Included$^{a}$  &   superflat   &   to make superflat   }
\startdata
1   &   $1-5$   &   FGK &   21  &   16  \\
2   &   $6-22$  &   FGK, SIM/TPF    &   30  &   21  \\
3   &   25  &   IRS, SIM/TPF    &   28  &   19  \\
4   &   $26-27$ &   IRS, FGK    &   27  &   16  \\
5   &   28  &   IRS, FGK    &   27  &   23  \\
6   &   29  &   IRS, FGK, SIM/TPF   &   35  &   30  \\
7   &   30  &   IRS &   25  &   22  \\
8   &   $31-38$ &   IRS &   20  &   19  \\
\enddata
\tablenotetext{a}{IRS refers to this paper, FGK refers to \citet{beichman06IRS}, and SIM/TPF refers to \citet{beichman06SIM}}
\end{deluxetable}

\clearpage
\begin{deluxetable}{l|lcc|l}									
\tablewidth{0pt} \tabletypesize{\scriptsize} \tablecaption{Stars with Two Observations\label{pairtable}}									
\tablehead{									
Star	&	AOR	&	Program$^a$	&	$\chi$32	&	Comments	}
\startdata									
HD 115617	&	12718080	&	FGK	&	3.7	&	Strong excess	\\	
	&	15998976	&	This paper	&	6.4	&		\\	
	&	coadd	&	This paper	&	5.3	&		\\	\hline
HD 117043	&	12718336	&	FGK	&	1.6	&	Weak excess	\\	
	&	16021248	&	This paper	&	4.3	&		\\	
	&	coadd	&	This paper	&	2.8	&		\\	\hline
HD 158633	&	10272256	&	SIM/TPF	&	3.7	&	Strong excess	\\	
	&	16030208	&	This paper	&	4.4	&		\\	
	&	coadd	&	This paper	&	4.4	&		\\	\hline
HD 185144	&	4024576	&	FGK	&	0.7	&	No excess	\\	
	&	15999744	&	This paper	&	0.8	&		\\	
	&	coadd	&	This paper	&	0.8	&		\\	\hline
HD 190406	&	13473536	&	SIM/TPF	&	2.0	&	No excess	\\	
	&	16001792	&	This paper	&	0.0	&		\\	
	&	coadd	&	This paper	&	1.0	&		\\	\hline
HD 199260	&	10272000	&	SIM/TPF	&	5.7	&	Strong excess	\\	
	&	16003840	&	This paper	&	6.1	&		\\	
	&	coadd	&	This paper	&	6.3	&		\\	\hline
HD 222237	&	13473280	&	SIM/TPF	&	-0.2	&	No excess	\\	
	&	16002304	&	This paper	&	0.1	&		\\	
	&	coadd	&	This paper	&	-0.1	&		\\	
\enddata										
\tablenotetext{a}{FGK refers to \citet{beichman06IRS} and SIM/TPF refers to \citet{beichman06SIM}}									
\end{deluxetable}

\clearpage
\begin{deluxetable}{l|cccc|cccc}																													
\tablewidth{0pt} \tabletypesize{\scriptsize} \tablecaption{IRS Data\label{IRStable}}																													
\tablehead{																													
				&	\multicolumn{4}{c}{8.5--12 $\micron$} 											&	\multicolumn{4}{c}{30--34 $\micron$} 											\\	\cline{2-5}\cline{6-9}
Star				&	Excess			&	Fractional	&		&	L$_{dust}$/L$_*^{a}$			&	Excess			&	Fractional 	&		&	L$_{dust}$/L$_*^{a}$			\\	
				&	(mJy)			&	Excess	&	$\chi10$	&	($\times$10$^{-5}$)			&	(mJy)			&	Excess	&	$\chi32$	&	($\times$10$^{-5}$)			}	
\startdata																													
HD 870$^{b}$			&	2.3	$\pm$	0.4	&	0.009	&	0.8	&	$<$ 	12		&	2.2	$\pm$	0.2	&	0.082	&	3.8	&		1.3		\\	
HD 1461$^{b}$			&	4.0	$\pm$	0.8	&	0.010	&	1.0	&	$<$ 	8.6		&	7.0	$\pm$	1.0	&	0.162	&	5.1	&		1.8		\\	
HD 1581$^{c}$			&	16.5	$\pm$	1.2	&	0.005	&	0.5	&	$<$ 	8.3		&	2.0	$\pm$	1.4	&	0.006	&	0.3	&	$<$ 	0.9		\\	
HD 3765				&	0.1	$\pm$	0.5	&	0.000	&	0.0	&	$<$ 	14		&	-0.4	$\pm$	0.4	&	-0.014	&	-0.6	&	$<$ 	1.5		\\	
HD 4308				&	0.0	$\pm$	0.5	&	0.000	&	0.0	&	$<$ 	9.9		&	-0.2	$\pm$	0.2	&	-0.004	&	-0.2	&	$<$ 	1.1		\\	
HD 4813$^{d}$			&	2.6	$\pm$	0.8	&	0.002	&	0.2	&	$<$ 	7.5		&	-0.5	$\pm$	0.8	&	-0.004	&	-0.2	&	$<$ 	0.8		\\	
HD 5133				&	1.9	$\pm$	0.7	&	0.004	&	0.4	&	$<$ 	15		&	0.5	$\pm$	0.5	&	0.013	&	0.6	&	$<$ 	1.6		\\	
HD 7439				&	-7.0	$\pm$	0.7	&	-0.009	&	-0.9	&	$<$ 	6.8		&	-0.7	$\pm$	0.8	&	-0.007	&	-0.3	&	$<$ 	0.7		\\
\enddata																										
\tablenotetext{a}{|Limits on L$_{dust}$/L$_{*}$ from using Equation~\ref{dusteq} and 3 times the dispersion in the fractional excess of the whole population}																										
\tablenotetext{b}{Star has public MIPS 24 and 70 $\micron$ data that was reduced as described in $\S$ \ref{MIPSphot}}																										
\tablenotetext{c}{Star has MIPS 24 and 70 $\micron$ data from the FGK survey \citep{beichman06IRS}}																										
\tablenotetext{d}{Star has MIPS 24 and 70 $\micron$ data from the SIM/TPF survey \citep{beichman06SIM}}																										
\tablenotetext{e}{Data has significant pointing errors or a close companion so IRS data may not be accurate.}	
					
\tablenotetext{f}{Star has MIPS 24 and 70 $\micron$ data from \citet{plavchan08}}	

\tablenotetext{g}{Star has MIPS 24 and 70 $\micron$ data from \citet{trilling08}}																										
\tablenotetext{h}{IRS data has been co-added with IRS data from the FGK or SIM/TPF program (see Table~\ref{pairtable}).}																																												\tablenotetext{i}{Average values of L$_{dust}$/L$_*$ (3$\sigma$ limits or detections) after 2$\sigma$ rejection of outliers.}																										
\end{deluxetable}

\clearpage
\begin{deluxetable}{l|ccc|ccccc|l}																						
\tablewidth{0pt} \tabletypesize{\scriptsize} \tablecaption{Stars with IRS and/or MIPS Excesses\label{MIPStable}}																						
\tablehead{																						
Star	&	\multicolumn{3}{c}{IRS (30--34 $\micron$)}					&	\multicolumn{5}{c}{MIPS 70 $\micron$} 											&	MIPS	\\	\cline{2-4} \cline{5-9}
	&	Fractional	&		&		&	F$_{\nu}$ 			&	Excess	&	Fractional	&		&		&	data	\\	
	&	Excess	&	$\chi$32	&	Excess?	&	(mJy)			&	(mJy)	&	Excess	&	$\chi$70	&	Excess?	&	source$^a$	}	
\startdata																						
HD 870	&	0.08	&	3.8	&	yes	&	21	$\pm$	4	&	15	&	2.6	&	3.9	&	yes	&	This paper$^b$	\\	
HD 1461	&	0.16	&	5.1	&	yes	&	61	$\pm$	6	&	52	&	5.7	&	8.6	&	yes	&	This paper$^b$	\\	
HD 10647	&	1.20	&	21.7	&	yes	&	878	$\pm$	8$^c$	&	862	&	52.1	&	94.1	&	yes	&	\citet{bryden09}	\\	
HD 38858	&	0.19	&	6.8	&	yes	&	190	$\pm$	9$^c$	&	175	&	11.7	&	17.2	&	yes	&	\citet{bryden09}	\\	
HD 40136	&	0.16	&	7.3	&	yes	&	95	$\pm$	5	&	39	&	0.7	&	6.4	&	yes	&	\citet{beichman06SIM}	\\	
HD 45184	&	0.38	&	12.8	&	yes	&	119	$\pm$	7	&	110	&	12.1	&	15.9	&	yes	&	This paper$^b$	\\	
HD 76653	&	0.20	&	8.1	&	yes	&	38	$\pm$	11	&	26	&	2.1	&	4.3	&	yes	&	\citet{beichman06SIM}	\\	
HD 90089	&	0.01	&	0.7	&	no	&	41	$\pm$	3	&	24	&	1.5	&	8.3	&	yes	&	\citet{beichman06SIM}	\\
\enddata																					
\tablecomments{There are 60 additional stars with MIPS data, but no excesses. These are noted in Table~\ref{IRStable}}																																				
\tablenotetext{a}{MIPS data was reduced as described in \S\ref{MIPSphot}}

\tablenotetext{b}{Data from {\it Spitzer} GO Program 30490, D. Koerner, P.I.}

\tablenotetext{c}{Flux from a model fit to the resolved disk, not aperture photometry}													
\tablenotetext{d}{Data from {\it Spitzer} GTO Program 50150, G. Rieke, P.I.}	

\tablenotetext{e}{This star is located in a particularly noisy field, so this cannot be considered an excess}												
\end{deluxetable}																																																	
\begin{deluxetable}{l|ccl|ccl}														
\tablewidth{0pt} \tabletypesize{\scriptsize} \setlength{\tabcolsep}{0.05in} \tablecaption{Stars with Previous IRS Observations\label{rejecttable}}														

\clearpage
\tablehead{														
Star	&	\multicolumn{3}{c}{IRS Data} & \multicolumn{3}{c}{MIPS Data}  \\	
\cline{2-4} \cline{5-7}
&	8.5-12 $\micron$	&	30-34 $\micron$	&	IRS data	
&	24 $\micron$	&	70 $\micron$	&	MIPS data 	\\
&	excess?	&	excess?	&	source:	
&	excess?	&	excess?	&	source:	}	 
\startdata														
HD 7570	
&	no	&	yes	&	\citet{beichman06IRS}	
&	weak	&	weak	&	\citet{bryden06}	\\
HD 69830	
&	yes	&	yes	&	\citet{beichman06IRS}	
&	yes	&	weak	&	\citet{bryden06}	\\
HD 72905 	
&	weak	&	yes	&	\citet{beichman06IRS}	
&	weak	&	yes	&	\citet{bryden06}	\\
HD 76151 	
&	no	&	yes	&	\citet{beichman06IRS}	
&	no	&	yes	&	\citet{bryden06}	\\
HD 109085	
&	yes	&	yes	&	Examination of archive data$^b$, \citet{chen06}	
&	yes	&	yes	&	\citet{beichman06SIM}	\\
HD 128311	
&	no	&	no	&	\citet{beichman06IRS}	
&	no	&	yes	&	\citet{beichman05FGK}	\\
HD 151044	
&	no	&	yes	&	Examination of archive data$^c$	
&	weak	&	yes	&	This paper$^{a,c}$	\\
HD 206860	
&	no	&	yes	&	\citet{beichman06IRS}	
&	no	&	yes	&	\citet{bryden06}	\\
\enddata													
\tablenotetext{a}{Public MIPS 24 and 70 $\micron$ data was reduced as described in \S\ref{MIPSphot}}												
\tablenotetext{b}{Data from {\it Spitzer} GTO Program 2, J. Houck, P.I.}													
\tablenotetext{c}{Data from {\it Spitzer} GO Program 3401, P. Abraham, P.I.}													
\tablenotetext{d}{Data from {\it Spitzer} GO Program 20065, G. Bryden, P.I.}													
\tablenotetext{e}{Data from {\it Spitzer} Legacy Program 148, M. Meyer, P.I.}													
\tablenotetext{f}{Data from {\it Spitzer} GO Program 30490, D. Koerner, P.I.}													
\end{deluxetable}

\clearpage
\begin{deluxetable}{lcc|cc|ccccc}																															
\tablewidth{0pt} \tabletypesize{\scriptsize} \rotate \tablecaption{Large Dust Models\label{dusttable}}																															
\tablehead{																															
	&		&	Spectral	&	$R_1-R_2$			&	$T_1-T_2$			&	Reduced		&	Optical Depth $\tau$			&	M$_{grain}$$^a$			&		&	$\ld$$^b$				\\
Star	&	Excess	&	Type	&	(AU)			&	(K)			&	$\chi$$^2$		&	(10 $\micron$)			&	(M$_{\earth}$)			&	(M$_{\earth}$ $<$ 10 km)	&	($10^{-7}$)				}
\startdata																															
HD 870	&	IRS and MIPS 70$\micron$	&	K0V	&	12	$-$	12	&	58	$-$	57	&	0.8$^{c}$	&	1.5	$\times$	10$^{-3}$	&	7.8	$\times$	10$^{-6}$	&	0.2	&	245	$\pm$	91		\\
HD 1461	&	IRS and MIPS 70$\micron$	&	G0V	&	19	$-$	26	&	62	$-$	54	&	0.9$^{c}$	&	2.7	$\times$	10$^{-4}$	&	3.4	$\times$	10$^{-5}$	&	1.06	&	343	$\pm$	45		\\
HD 10647	&	IRS and MIPS 70$\micron$	&	F9V	&	16	$-$	29	&	69	$-$	52	&	5.7$^{d}$	&	10.0	$\times$	10$^{-4}$	&	2.4	$\times$	10$^{-4}$	&	7.7	&	2627	$\pm$	118$^{e}$		\\
HD 38858	&	IRS and MIPS 70$\micron$	&	G4V	&	12	$-$	29	&	69	$-$	44	&	1.0$^{c}$	&	1.7	$\times$	10$^{-4}$	&	5.3	$\times$	10$^{-5}$	&	1.7	&	677	$\pm$	141		\\
HD 40136	&	IRS and MIPS 70$\micron$	&	F1V	&	1	$-$	16	&	353	$-$	92	&	0.8$^{d}$	&	1.1	$\times$	10$^{-5}$	&	1.2	$\times$	10$^{-6}$	&	0.039	&	62	$\pm$	14		\\
HD 45184	&	IRS and MIPS 70$\micron$	&	G2IV	&	11	$-$	27	&	77	$-$	49	&	1.2$^{d}$	&	2.0	$\times$	10$^{-4}$	&	5.3	$\times$	10$^{-5}$	&	1.68	&	810	$\pm$	89		\\
HD 76653	&	IRS and weak MIPS 70$\micron$	&	F6V	&	16	$-$	18	&	77	$-$	73	&	0.9$^{c}$	&	4.7	$\times$	10$^{-4}$	&	1.0	$\times$	10$^{-5}$	&	0.33	&	164	$\pm$	47		\\
HD 90089	&	IRS limit and MIPS 70$\micron$	&	F2V	&	$>$15			&	$<$70			&	0.8$^{c}$	&	\nodata			&	\nodata	&	\nodata	&	167	$\pm$	21		\\
\enddata																															
\tablenotetext{a}{Calculated assuming 10$\micron$ dust grains.}

\tablenotetext{b}{See note in \S\ref{simplemodel} for how $\ld$ is calculated.}																														
\tablenotetext{c}{$\lambda$ $>$ 21 $\micron$, 74 dof.}																															
\tablenotetext{d}{$\lambda$ $>$ 14 $\micron$, 152 dof.}	

\tablenotetext{e}{\citet{tanner08} calculate $\ld$ = 2770 $\times$ 10$^{-7}$ including the MIPS 160 $\micron$ datapoint.}		

\tablenotetext{f}{\citet{tanner08} calculate $\ld$ = 259 $\times$ 10$^{-7}$ including the MIPS 160 $\micron$ datapoint.}																										
\tablenotetext{g}{This star is located in a particularly noisy field, so the model is fit to the IRS data only, and $\ld$ is calculated using data out to 35 $\micron$ only.}																															
\end{deluxetable}

\clearpage

%%%%%%%%%%%%%%%%%%%%%%%%%%%%%%%%%%%%%%%%%%%%%%%%%%%%%%%%%%%%%%%%%%%%
\clearpage
%%%%%%%%%%%%%%%%%%%%%%%%%%%%%%%%%%%%%%%%%%%%%%%%%%%%%%%%%%%%%%%%%%%

%%%%%%%%%%%%%%%%%%%%%%%%%%%%%%%%%%%%%%%%%%%%%%%%%%%%%%%%%%%%%%%%%%%%
\clearpage
%%%%%%%%%%%%%%%%%%%%%%%%%%%%%%%%%%%%%%%%%%%%%%%%%%%%%%%%%%%%%%%%%%%%

%%%%%%%%%%%%%%%%%%%%%%%%%%%%%%%%%%%%%%%%%%%%%%%%%%%%%%%%%%%%%%%%%%%%
\clearpage
%%%%%%%%%%%%%%%%%%%%%%%%%%%%%%%%%%%%%%%%%%%%%%%%%%%%%%%%%%%%%%%%%%%%

%%%%%%%%%%%%%%%%%%%%%%%%%%%%%%%%%%%%%%%%%%%%%%%%%%%%%%%%%%%%%%%%%%%%
\clearpage
%%%%%%%%%%%%%%%%%%%%%%%%%%%%%%%%%%%%%%%%%%%%%%%%%%%%%%%%%%%%%%%%%%%%

%%%%%%%%%%%%%%%%%%%%%%%%%%%%%%%%%%%%%%%%%%%%%%%%%%%%%%%%%%%%%%%%%%%%
\clearpage
%%%%%%%%%%%%%%%%%%%%%%%%%%%%%%%%%%%%%%%%%%%%%%%%%%%%%%%%%%%%%%%%%%%%

%%%%%%%%%%%%%%%%%%%%%%%%%%%%%%%%%%%%%%%%%%%%%%%%%%%%%%%%%%%%%%%%%%%%
\clearpage
%%%%%%%%%%%%%%%%%%%%%%%%%%%%%%%%%%%%%%%%%%%%%%%%%%%%%%%%%%%%%%%%%%%%

%%%%%%%%%%%%%%%%%%%%%%%%%%%%%%%%%%%%%%%%%%%%%%%%%%%%%%%%%%%%%%%%%%%%
\clearpage
%%%%%%%%%%%%%%%%%%%%%%%%%%%%%%%%%%%%%%%%%%%%%%%%%%%%%%%%%%%%%%%%%%%%

%%%%%%%%%%%%%%%%%%%%%%%%%%%%%%%%%%%%%%%%%%%%%%%%%%%%%%%%%%%%%%%%%%%%
\clearpage\begin{figure}
\includegraphics[scale=0.7]{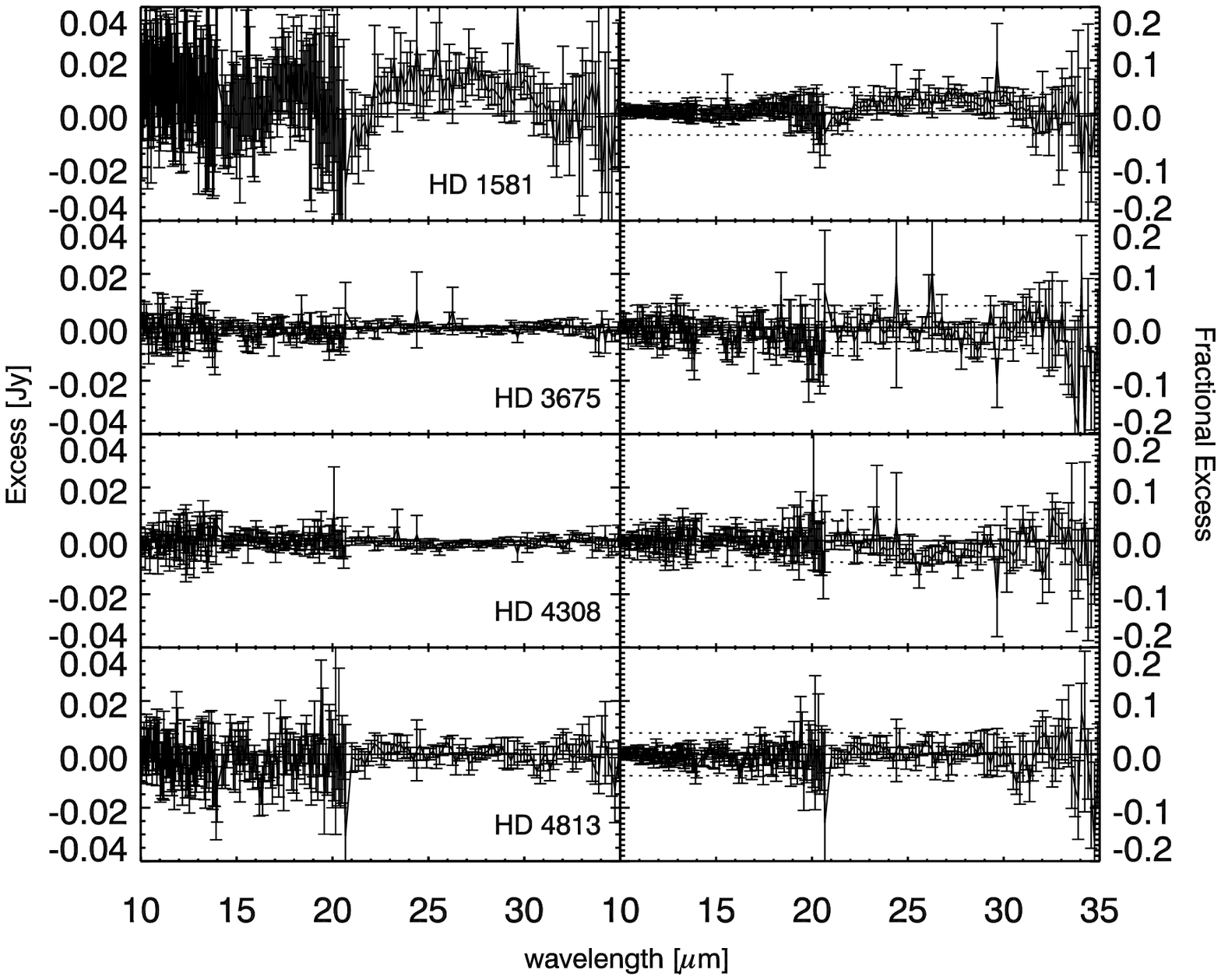}
\caption{IRS spectra for 4 stars with no excesses. The left hand plots show the excess in Jy relative to the photosphere after normalization with respect to the first 10 points of each module.  The right hand plots show the fractional amount of the excess relative to the photosphere after normalization. The dotted lines in the right panels show an estimate of the 2$\sigma$ dispersion in the deviations from the photospheric models. 
None of these stars have fractional excesses outside of these
2$\sigma$ limits.
\label{irsdataplot_noex}}
\end{figure}
%%%%%%%%%%%%%%%%%%%%%%%%%%%%%%%%%%%%%%%%%%%%%%%%%%%%%%%%%%%%%%%%%%%%

%%%%%%%%%%%%%%%%%%%%%%%%%%%%%%%%%%%%%%%%%%%%%%%%%%%%%%%%%%%%%%%%%%%%
\clearpage\begin{figure}
\includegraphics[scale=0.7]{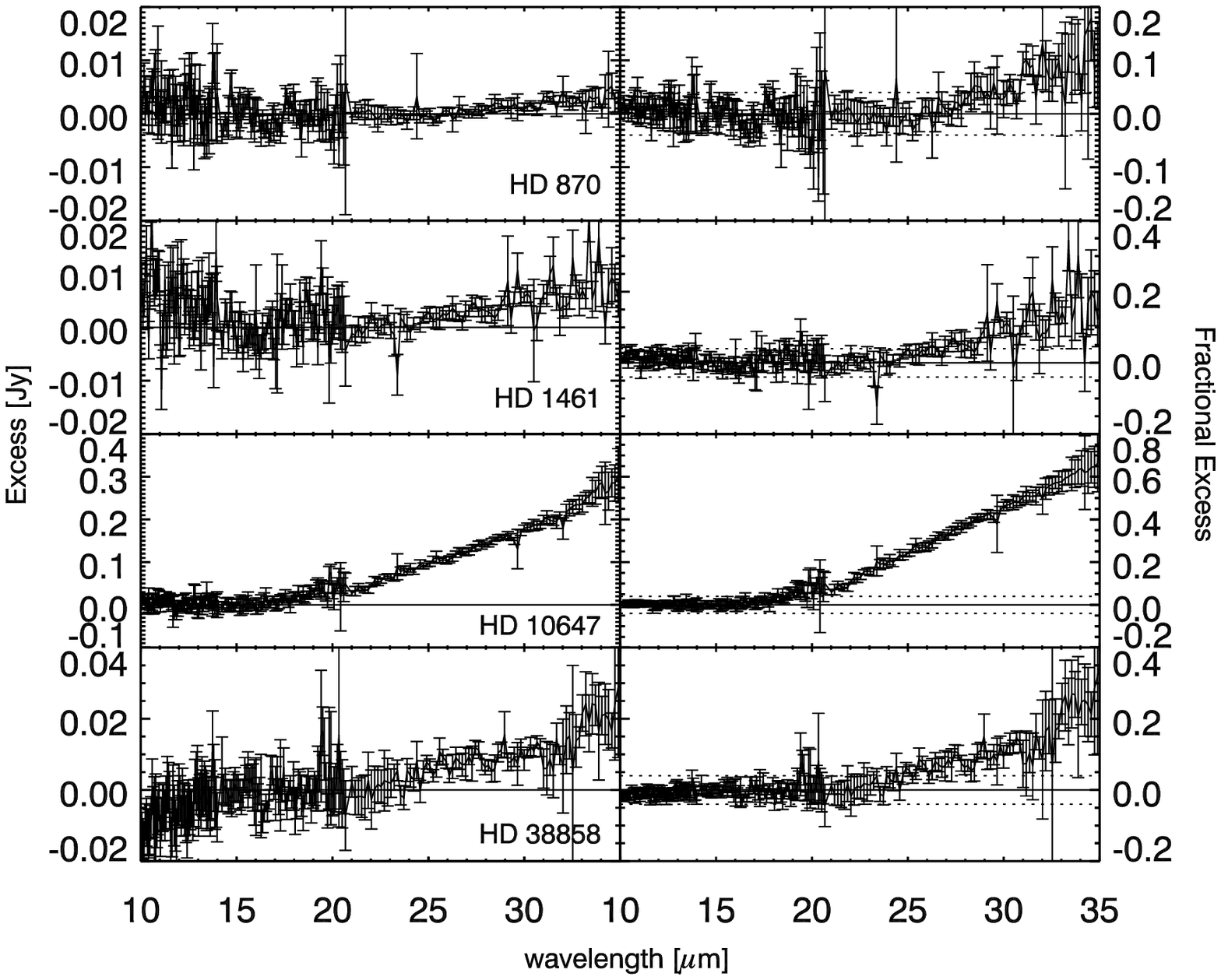}
\caption{IRS spectra for stars with significant excesses.  The left hand plots show the excess in Jy relative to the photosphere after normalization with respect to the first 10 points of each module.  The right hand plots show the fractional amount of the excess relative to the photosphere after normalization. The dotted lines in the right panels show an estimate of the 2$\sigma$ dispersion in the deviations from the photospheric models. 
All of these stars have fractional excesses that extend well above the 
2$\sigma$ threshold at longer wavelengths, and are significant at the 3$\sigma$ level in the 30--34 $\micron$ band.
\label{irsdataplot1}}
\end{figure}
%%%%%%%%%%%%%%%%%%%%%%%%%%%%%%%%%%%%%%%%%%%%%%%%%%%%%%%%%%%%%%%%%%%%

%%%%%%%%%%%%%%%%%%%%%%%%%%%%%%%%%%%%%%%%%%%%%%%%%%%%%%%%%%%%%%%%%%%%
\addtocounter{figure}{-1}
\clearpage\begin{figure}
\includegraphics[scale=0.7]{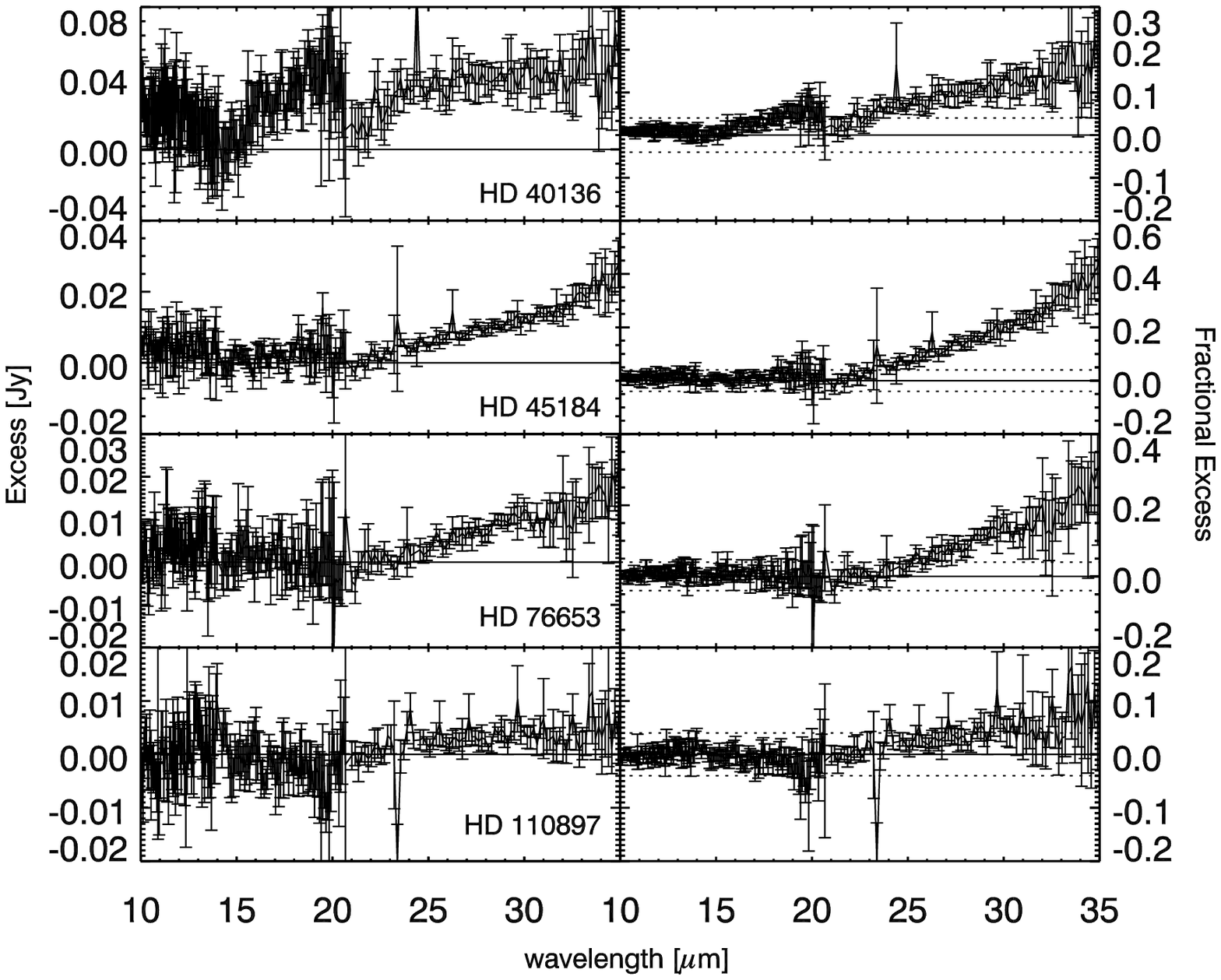}
\caption{continued}
\end{figure}
%%%%%%%%%%%%%%%%%%%%%%%%%%%%%%%%%%%%%%%%%%%%%%%%%%%%%%%%%%%%%%%%%%%%

%%%%%%%%%%%%%%%%%%%%%%%%%%%%%%%%%%%%%%%%%%%%%%%%%%%%%%%%%%%%%%%%%%%%
\addtocounter{figure}{-1}
\clearpage\begin{figure}
\includegraphics[scale=0.7]{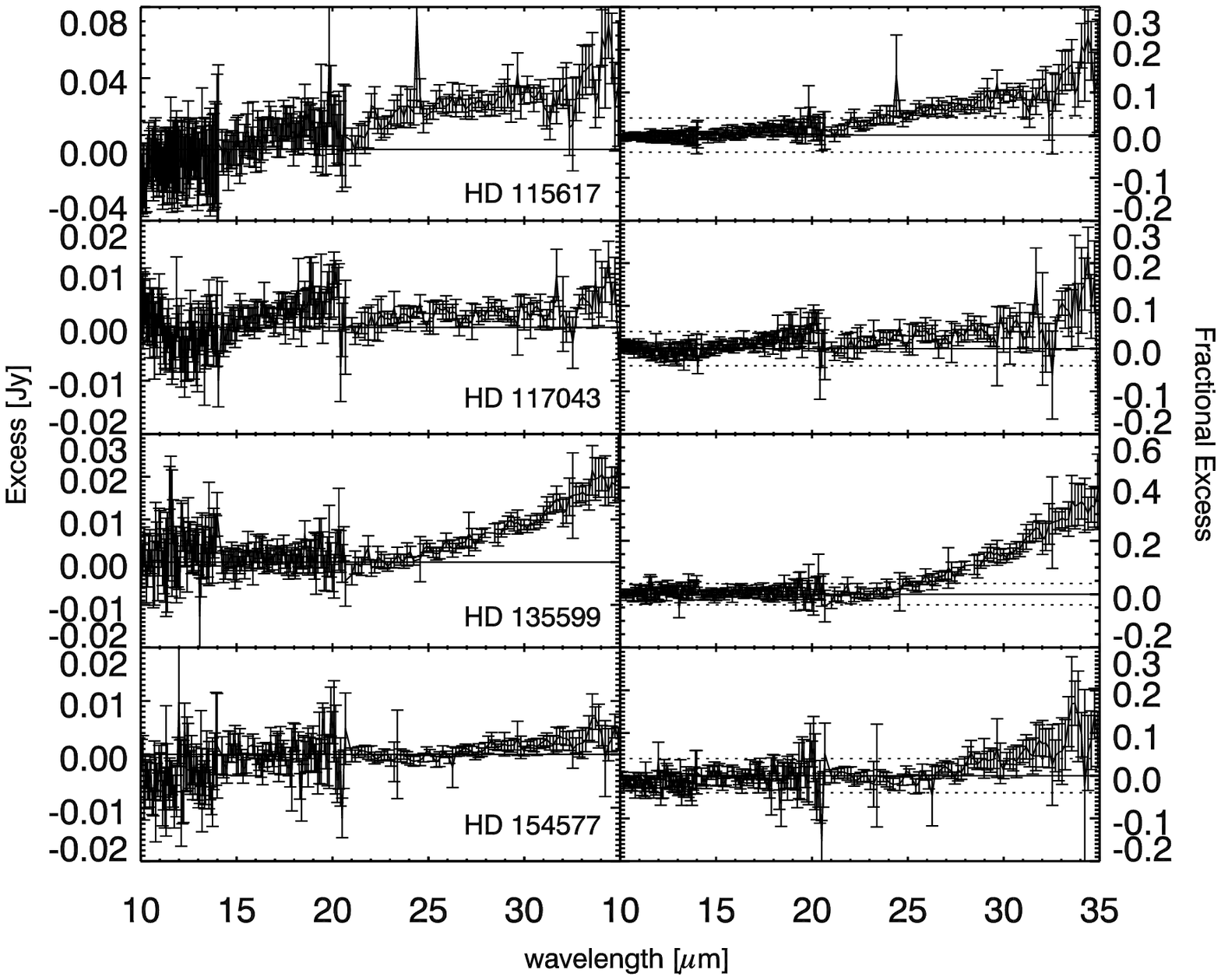}
\caption{continued}
\end{figure}
%%%%%%%%%%%%%%%%%%%%%%%%%%%%%%%%%%%%%%%%%%%%%%%%%%%%%%%%%%%%%%%%%%%%

%%%%%%%%%%%%%%%%%%%%%%%%%%%%%%%%%%%%%%%%%%%%%%%%%%%%%%%%%%%%%%%%%%%%
\addtocounter{figure}{-1}
\clearpage\begin{figure}
\includegraphics[scale=0.7]{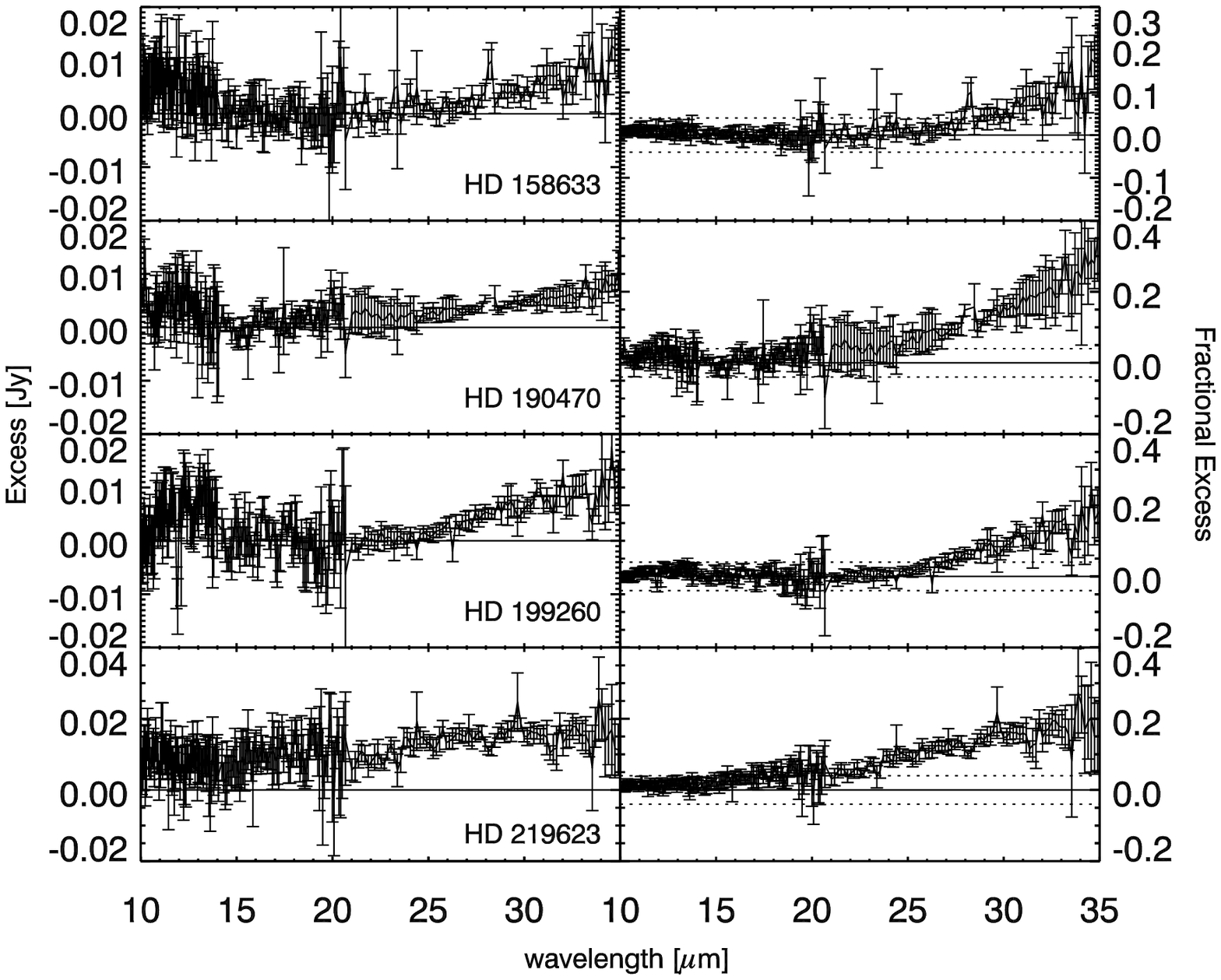}
\caption{continued}
\end{figure}
%%%%%%%%%%%%%%%%%%%%%%%%%%%%%%%%%%%%%%%%%%%%%%%%%%%%%%%%%%%%%%%%%%%%

%%%%%%%%%%%%%%%%%%%%%%%%%%%%%%%%%%%%%%%%%%%%%%%%%%%%%%%%%%%%%%%%%%%%
\clearpage\begin{figure}
\includegraphics[scale=0.7]{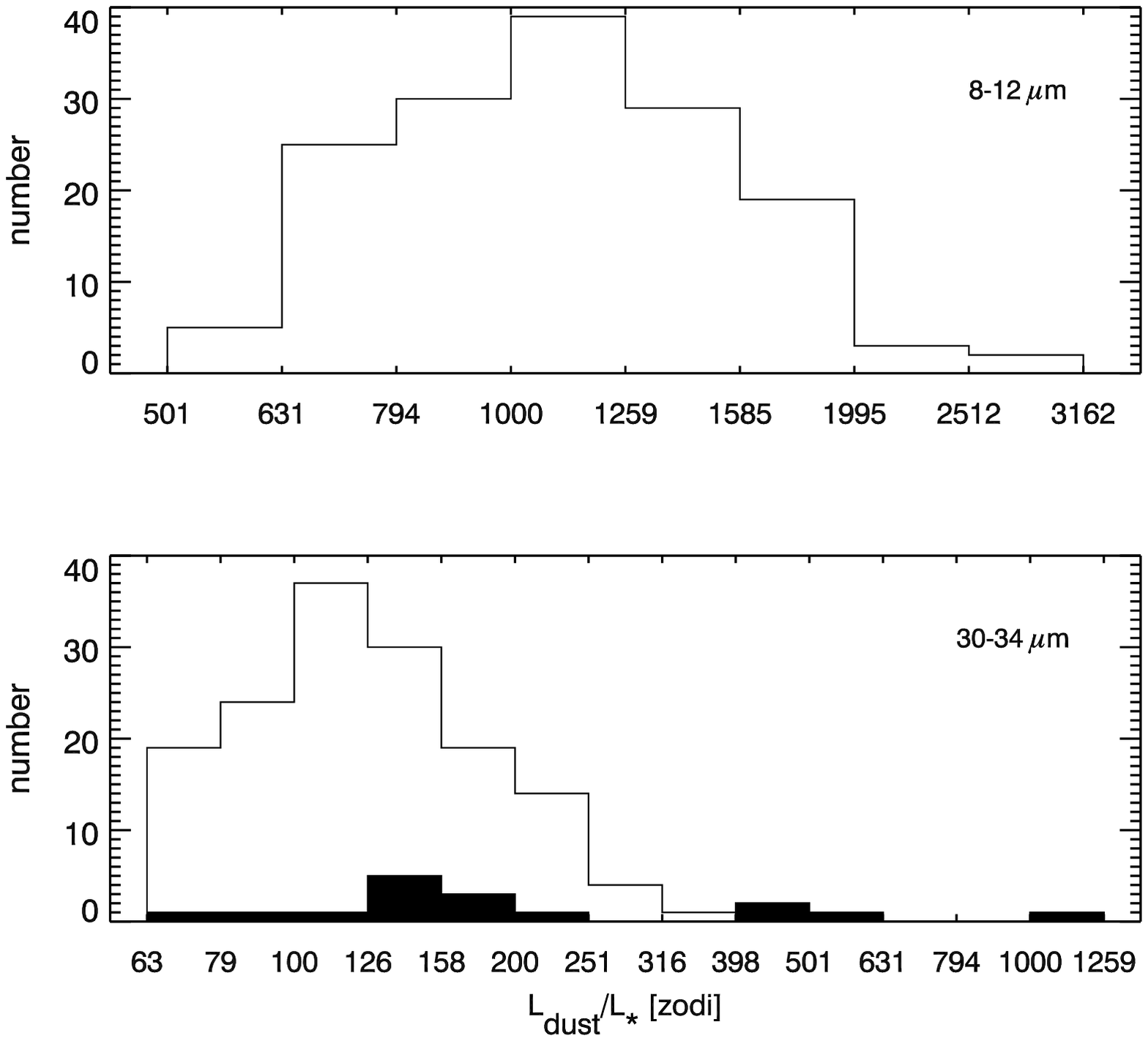}
\caption{Calculated values of $\ld$ for all sample stars in zodi (the value of $\ld$ in our solar system: $\sim$10$^{-7}$), using
the 8.5--12 $\micron$ photometric band in the upper panel
and using the 30--34 $\micron$ photometric band in the lower panel. All of the 8.5--12 $\micron$ measurements are upper limits based on
3$\sigma_{pop}$, as are most at 30-34 $\micron$; 
the 16 stars with significant excesses at 30--34 $\micron$ 
(filled bars) have $\ld$ calculated using Equation~\ref{dusteq}.
\label{LdLstarhist}}
\end{figure}
%%%%%%%%%%%%%%%%%%%%%%%%%%%%%%%%%%%%%%%%%%%%%%%%%%%%%%%%%%%%%%%%%%%%

%%%%%%%%%%%%%%%%%%%%%%%%%%%%%%%%%%%%%%%%%%%%%%%%%%%%%%%%%%%%%%%%%%%%
\begin{figure}
\begin{center}
\rotatebox{270}{\includegraphics[scale=0.5]{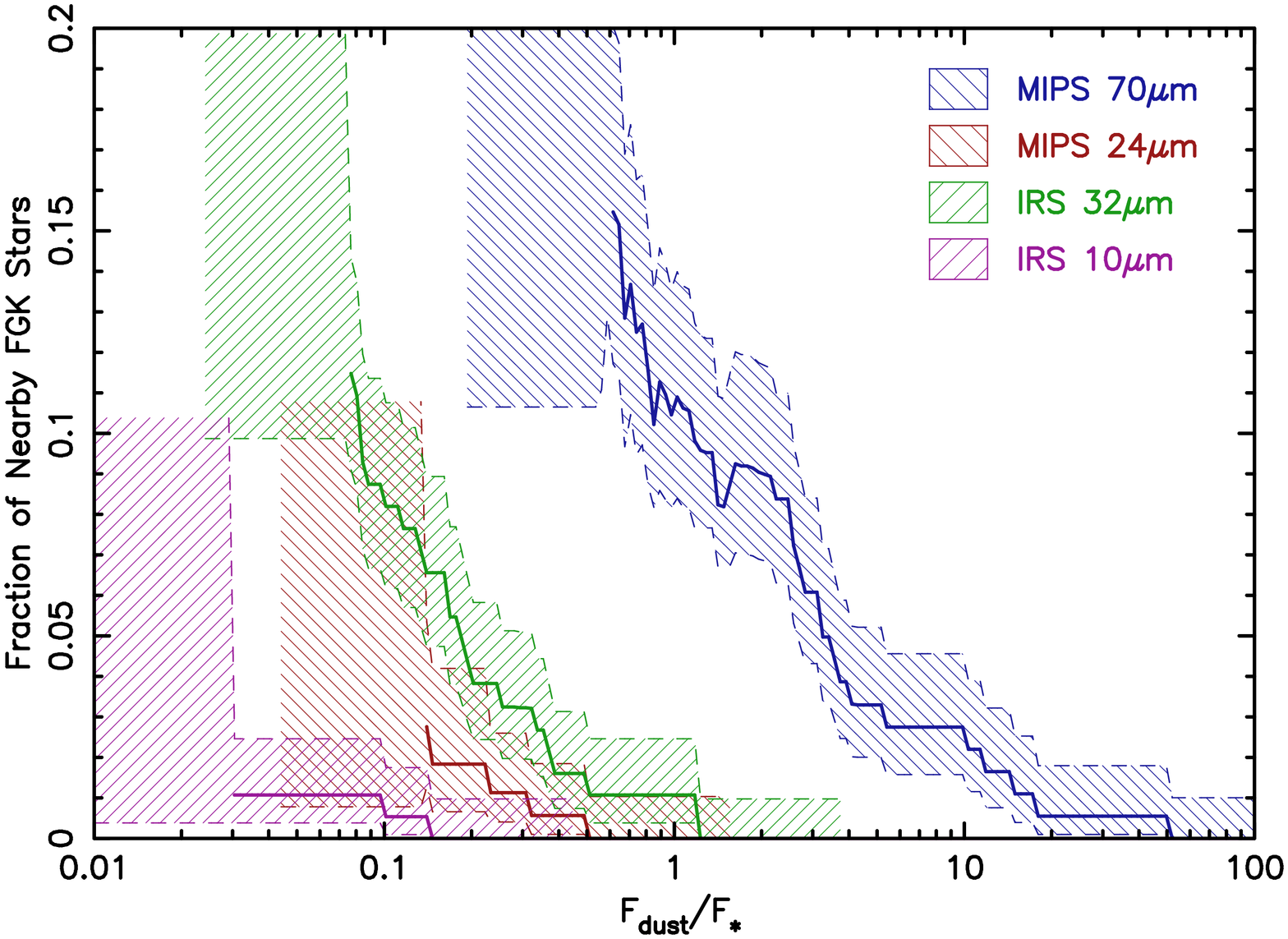} }
\end{center}
\caption{
\emph{Spitzer} detection rates of IR excess as a function of
the fractional dust flux, $\fd$.
The various \emph{Spitzer} instruments/wavelengths considered here are
indicated in the figure legend. 
For MIPS photometry, 182 stars with spectral types F5--K5 are
observed at 24 and 70 $\mu$m \citep{bryden06, beichman06SIM, trilling08}. 
For IRS spectra, 203 stars with spectral types F0--M0 are
observed from 10 through 32 $\mu$m (from this survey and the stars in
Table~\ref{rejecttable}). 
Uncertainties in the underlying distribution due to small number
statistics (shaded regions) are large below the detection limits of
each instrument/wavelength.
}\label{fluxrates}
\end{figure}
%%%%%%%%%%%%%%%%%%%%%%%%%%%%%%%%%%%%%%%%%%%%%%%%%%%%%%%%%%%%%%%%%%%%

%%%%%%%%%%%%%%%%%%%%%%%%%%%%%%%%%%%%%%%%%%%%%%%%%%%%%%%%%%%%%%%%%%%%
\clearpage\begin{figure}
\includegraphics[scale=0.7]{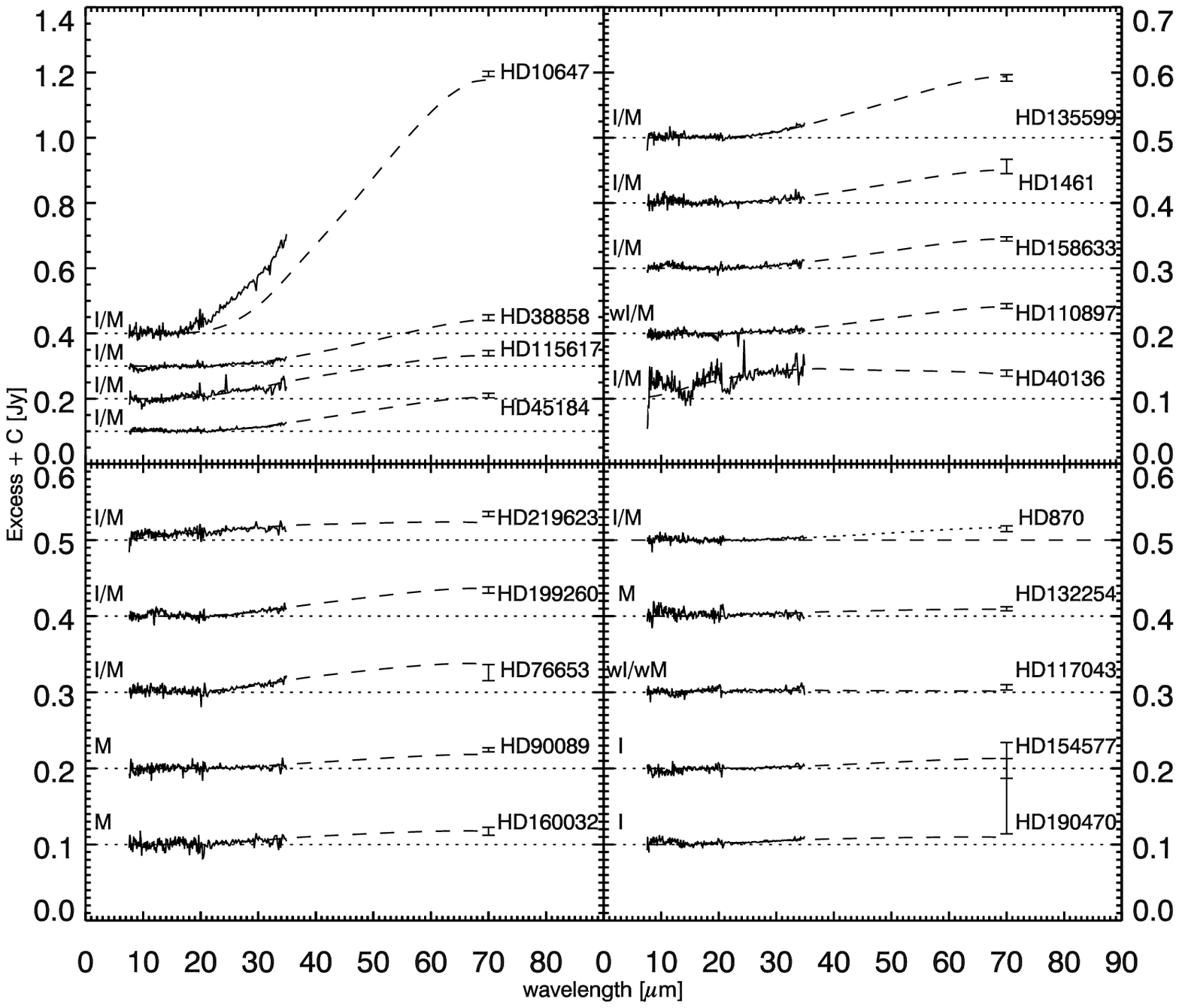}
\caption{Measured and modeled spectra (using the simple model, see $\S$ \ref{simplemodel}) for stars with IRS and/or 70
$\mu$m excesses. Refer to Figure~\ref{irsdataplot1} for error bars on the IRS data.  Measured spectra are shown with solid lines, modeled
spectra are shown with dashed lines. ``I'' indicates the star has an IRS excess, ``M'' indicates the star has a MIPS 70 $\micron$ excess, and a ``w'' indicates that the excess is
weak. 
\label{allexplotall}}
\end{figure}
%%%%%%%%%%%%%%%%%%%%%%%%%%%%%%%%%%%%%%%%%%%%%%%%%%%%%%%%%%%%%%%%%%%%

%%%%%%%%%%%%%%%%%%%%%%%%%%%%%%%%%%%%%%%%%%%%%%%%%%%%%%%%%%%%%%%%%%%%
\clearpage\begin{figure}
\includegraphics[scale=0.7]{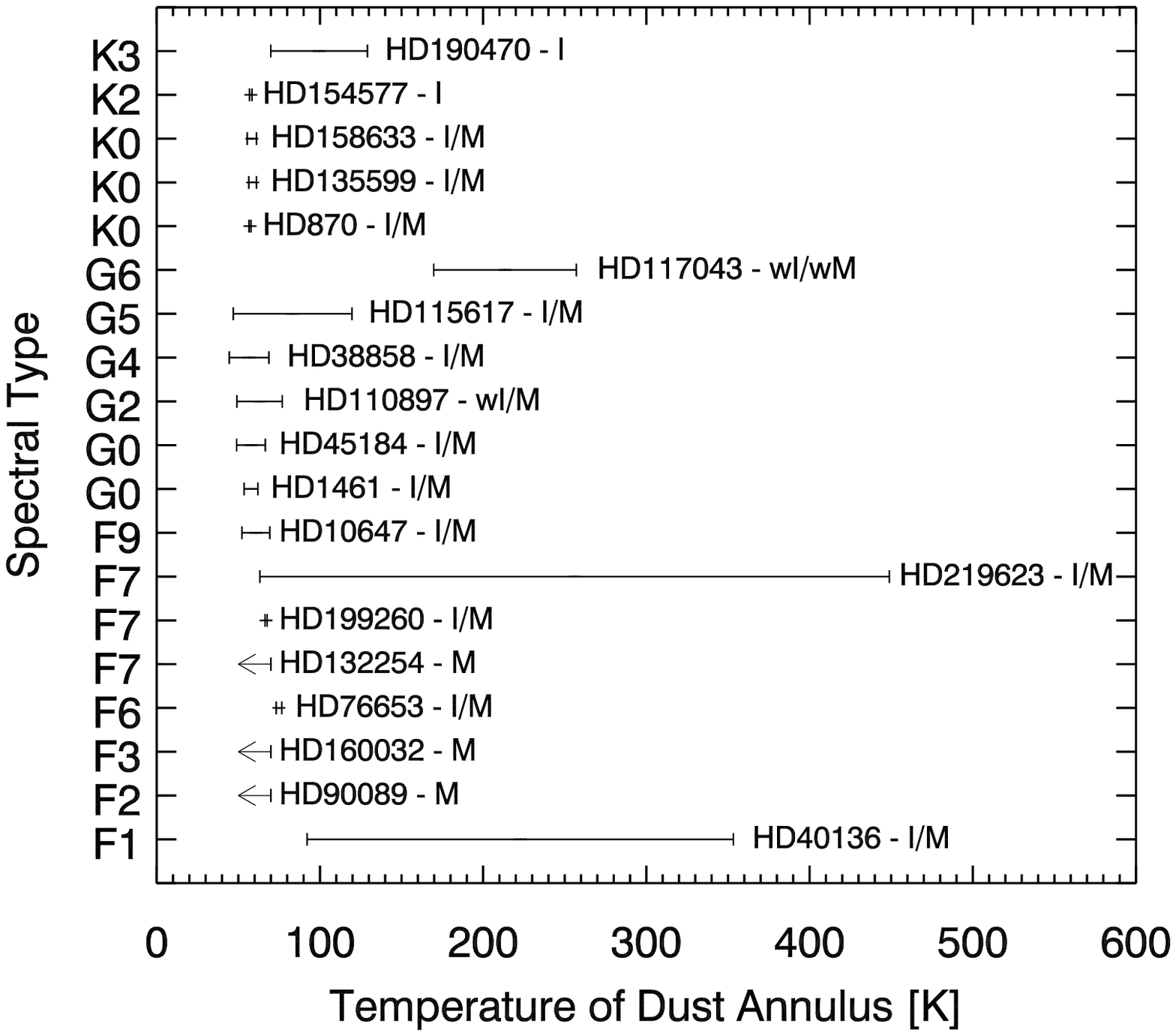}
\caption{Calculated dust temperatures based on model spectra
(Table~\ref{dusttable}; Figure~\ref{allexplotall}). Stars are arranged by spectral type. ``I''
indicates the star has an IRS excess, ``M'' indicates the star has a
MIPS 70 $\micron$ excess, and a ``w'' indicates that the excess is
weak. \label{t1t2}}
\end{figure}
%%%%%%%%%%%%%%%%%%%%%%%%%%%%%%%%%%%%%%%%%%%%%%%%%%%%%%%%%%%%%%%%%%%%

%%%%%%%%%%%%%%%%%%%%%%%%%%%%%%%%%%%%%%%%%%%%%%%%%%%%%%%%%%%%%%%%%%%%
\clearpage\begin{figure}
\includegraphics[scale=0.7]{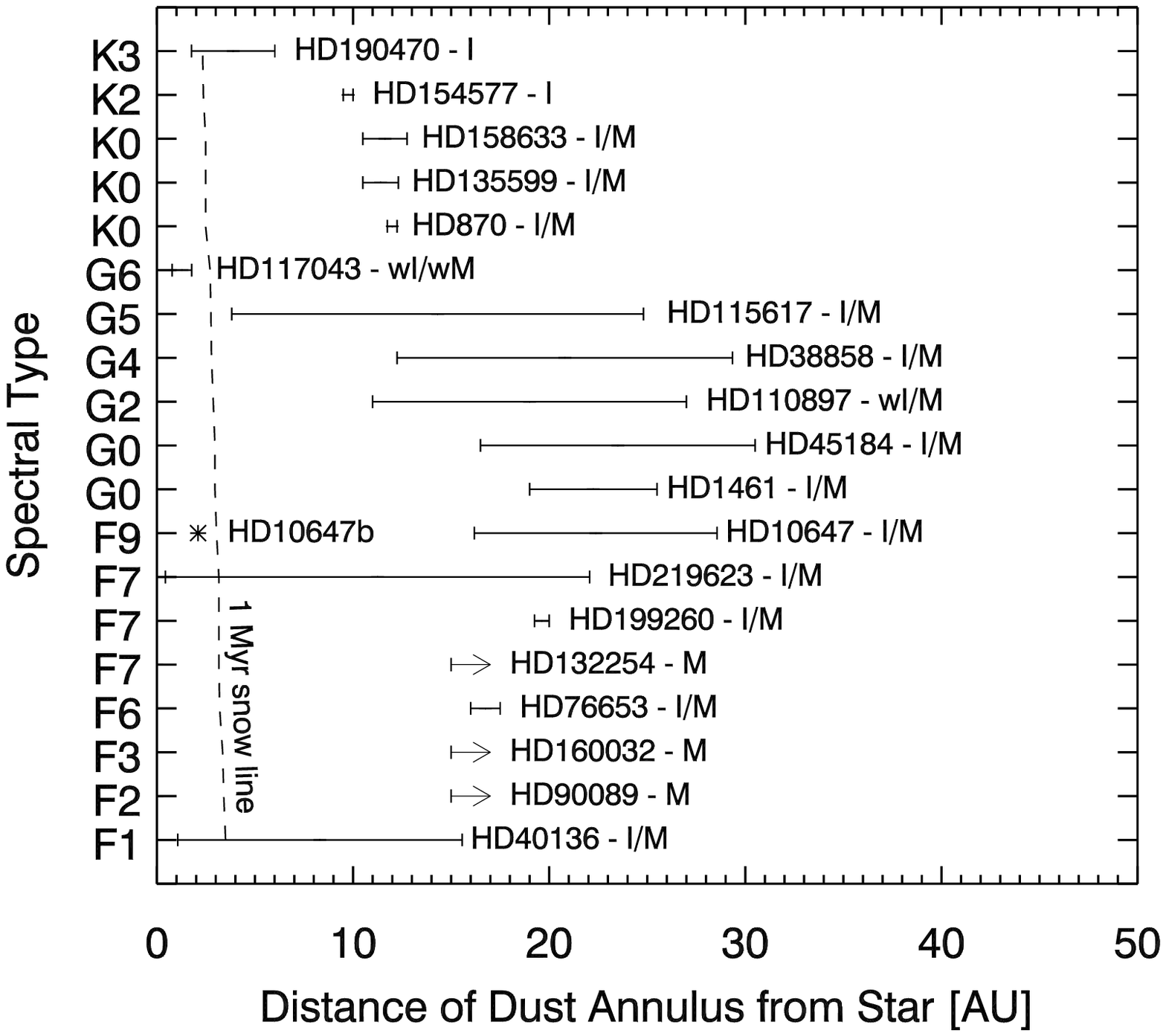}
\caption{Calculated dust radii based on model spectra
(Table~\ref{dusttable}; Figure~\ref{allexplotall}). Stars are arranged by spectral type. ``I''
indicates the star has an IRS excess, ``M'' indicates the star has a
MIPS 70 $\micron$ excess, and a ``w'' indicates that the excess is
weak. Also noted on this plot is the theoretical 1 Myr snow line \citep[based on][]{siess00} and the location of HD 10647's known planet. \label{r1r2}}
\end{figure}
%%%%%%%%%%%%%%%%%%%%%%%%%%%%%%%%%%%%%%%%%%%%%%%%%%%%%%%%%%%%%%%%%%%%

%%%%%%%%%%%%%%%%%%%%%%%%%%%%%%%%%%%%%%%%%%%%%%%%%%%%%%%%%%%%%%%%%%%%
\clearpage\begin{figure}
\includegraphics[scale=0.5]{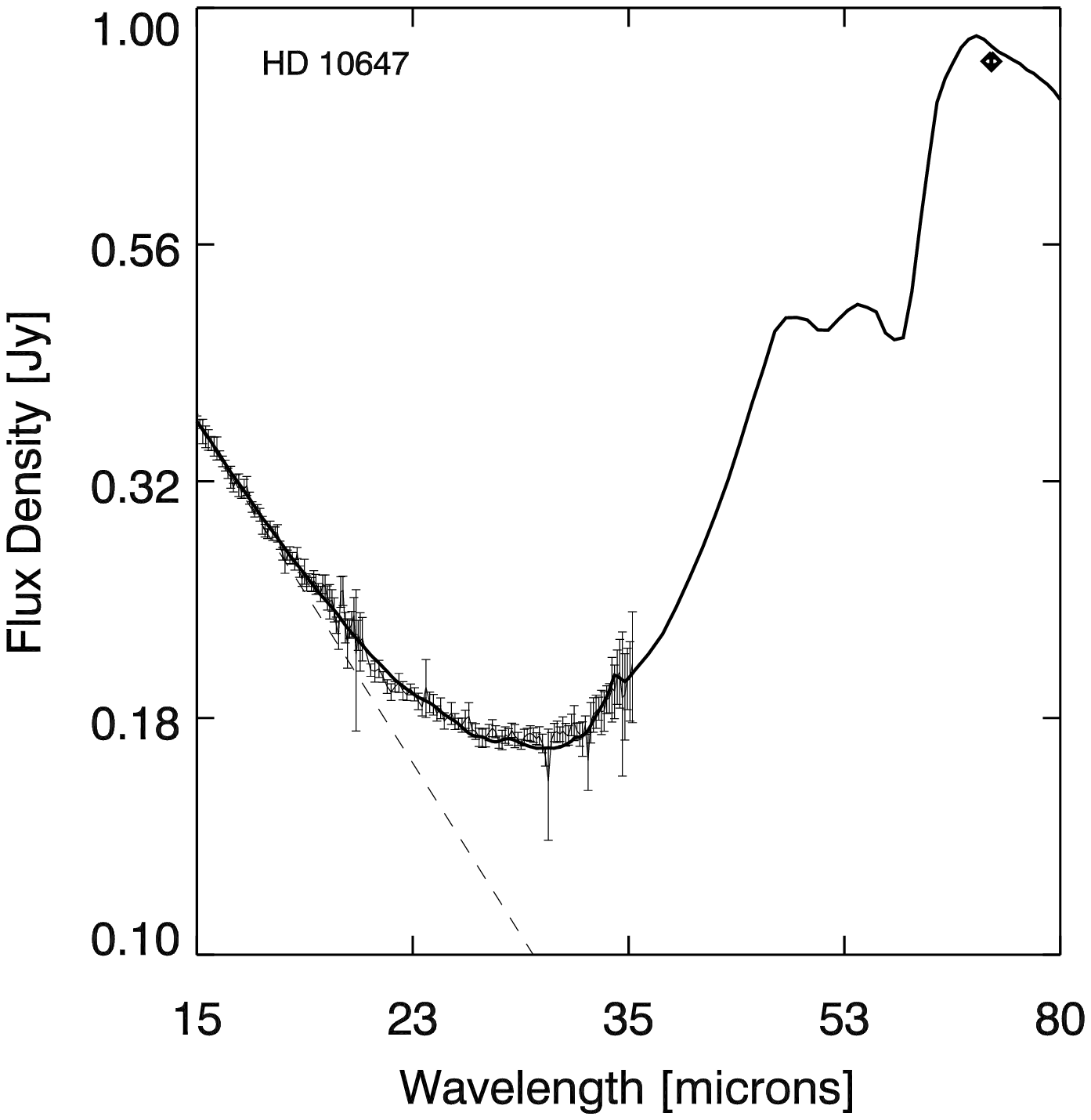}
\includegraphics[scale=0.5]{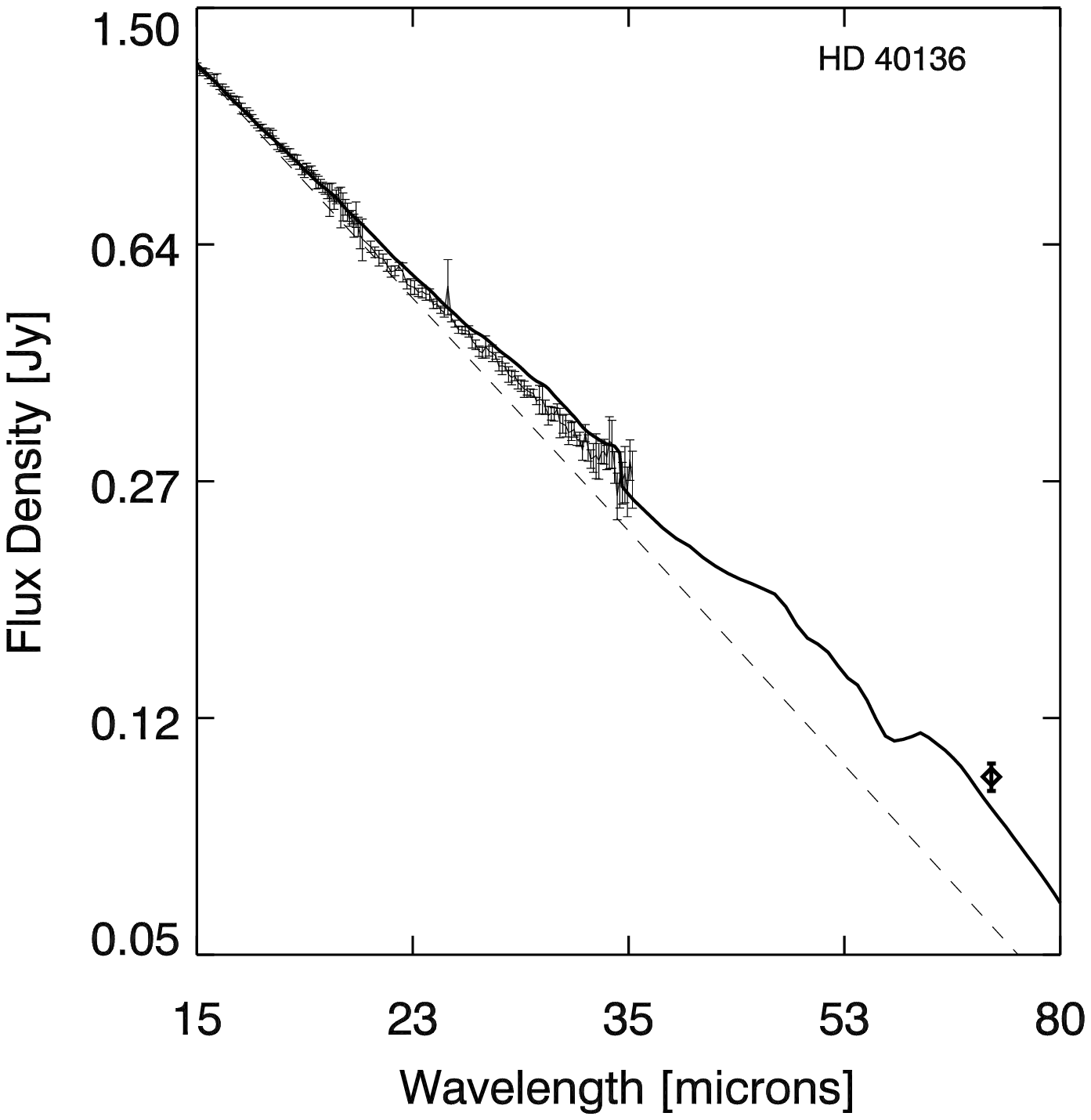}
\caption{The results of the mineralogical model discussed in \S\ref{othermodels1}, using a combination of water ice and silicates.  The dashed line shows the spectrum of the stellar photosphere, while the solid line shows the model.  Also shown are the IRS spectra and the MIPS 70 $\micron$ datapoint.  \label{caseyfigs}}
\end{figure}
%%%%%%%%%%%%%%%%%%%%%%%%%%%%%%%%%%%%%%%%%%%%%%%%%%%%%%%%%%%%%%%%%%%%

\clearpage
\appendix
\begin{center}
    {\bf APPENDIX: IRS Data for Stars with Excess}
\end{center}

\begin{deluxetable}{cccc}											
\tablewidth{0pt} \tabletypesize{\scriptsize} \tablecaption{HD 870\label{IRS870}}											
\tablehead{											
wavelength	&	F$_{\nu}$	&	Excess			&	Fractional			\\
($\micron$)	&	(Jy)	&	(Jy)			&	Excess			}
\startdata											
7.576	&	0.4784	&	-0.0020	$\pm$	0.0046	&	-0.0043	$\pm$	0.0097	\\
7.637	&	0.4721	&	-0.0006	$\pm$	0.0048	&	-0.0014	$\pm$	0.0102	\\
7.697	&	0.4692	&	0.0038	$\pm$	0.0052	&	0.0082	$\pm$	0.0110	\\
7.758	&	0.4583	&	-0.0004	$\pm$	0.0094	&	-0.0009	$\pm$	0.0204	\\
7.818	&	0.4536	&	0.0018	$\pm$	0.0054	&	0.0039	$\pm$	0.0118	\\
7.878	&	0.4400	&	-0.0052	$\pm$	0.0036	&	-0.0119	$\pm$	0.0081	\\
7.939	&	0.4430	&	0.0043	$\pm$	0.0051	&	0.0098	$\pm$	0.0114	\\
7.999	&	0.4382	&	0.0061	$\pm$	0.0048	&	0.0138	$\pm$	0.0109	\\
8.060	&	0.4201	&	-0.0053	$\pm$	0.0042	&	-0.0126	$\pm$	0.0100	\\
\enddata											
\end{deluxetable}

\end{document}